\documentclass{aa}  
\usepackage[]{amsmath}  
\usepackage{graphicx}
\usepackage{color}
\usepackage{txfonts}
\usepackage{url}
\usepackage{natbib}         
\bibpunct{(}{)}{;}{a}{}{,}  
\newcommand{\vc}[1]{\mbox{\boldmath{$#1$}}}  

\begin{document} 

\title{Spontaneous concentrations of solids 
through \\ two-way drag forces between gas and sedimenting particles
}
\titlerunning{Spontaneous concentrations of solids through two-way drag
forces}

\author{
  M. Lambrechts \inst{1,2} \and 
  A. Johansen\inst{1}    \and
  H. L. Capelo\inst{3}   \and
  J. Blum\inst{4}        \and
  E. Bodenschatz\inst{3} 
} 
\institute{
Lund Observatory, Department of Astronomy and Theoretical
Physics, Lund University, Box 43, 22100 Lund, Sweden\\
\email{michiel@astro.lu.se}
\and
Laboratoire Lagrange, UMR7293, Universit\'e de Nice Sophia-Antipolis,
Observatoire de la C\^ote d'Azur, Boulevard de l'Observatoire, 06304 Nice Cedex 4, France
\and
Max Planck Institute for Dynamics and Self-Organization, (MPIDS), 37077
G\"ottingen, Germany
\and
Institut für Geophysik und extraterrestrische Physik, Technische Universität
Braunschweig, Mendelssohnstr.\,3, 38106 Braunschweig, Germany
}

\date{Received --- ; accepted ---}

\abstract{
The behaviour of sedimenting particles depends on the dust-to-gas ratio of the
fluid.
Linear stability analysis shows that solids settling in the Epstein drag regime
would remain homogeneously distributed in non-rotating incompressible
fluids, even when dust-to-gas ratios reach unity.
However, the non-linear evolution has not been probed before.
Here, we present numerical calculations indicating that in a particle-dense
mixture solids spontaneously mix out of the fluid and form swarms overdense in
particles by at least a factor $10$.
The instability is caused by mass-loaded regions locally breaking the
equilibrium background stratification.
The driving mechanism depends on non-linear perturbations of the background flow
and shares some similarity to the streaming instability in accretion discs.
The resulting particle-rich swarms may stimulate particle growth by coagulation. 
In the context of protoplanetary discs, the instability could be relevant for
aiding small particles to settle to the midplane in the outer disc.
Inside the gas envelopes of protoplanets, enhanced settling may lead to a
reduced dust opacity, which facilitates the contraction of the envelope.
We show that the relevant physical set up can be recreated in a laboratory
setting. This will allow our numerical calculations to be investigated
experimentally in the future.
}

\keywords{
  Hydrodynamics,
  Instabilities,
  Turbulence,
  Methods: numerical,
  Planets and satellites: formation, 
  Protoplanetary disks }

\maketitle

\section{Introduction}

The study of gas drag on mm to dm-sized particles (pebbles) is essential to understand the formation of planets. 
Vertical sedimentation due to drag on small particles in the protoplanetary
disc is necessary for the creation of a dense midplane of solids from which
larger objects can grow \citep{Youdin_2007b}. 
Conversely, the same drag force is also responsible for the rapid radial
migration of pebbles in the midplane \citep[on $100$\,yr time scales in the
terrestrial region,][]{Weidenschilling_1977}.
This is the main barrier for continued growth to larger than cm to m sizes by
collisions \citep{Brauer_2008,Birnstiel_2012}, 
unless particles can have extremely low internal densities
\citep[][]{Kataoka_2013,Krijt_2015}. 

The radial drift hurdle can be avoided through two mechanisms that also
critically rely on gas drag.
Firstly, pebbles can be concentrated hydrodynamically, so that the resulting
clouds collapse gravitationally to planetesimals of $\sim$$100$\,km in size
\citep[for recent reviews on different planetesimal formation models, see
][]{Johansen_2014, Chiang_2010}.
Secondly, large planetesimals can accrete the remaining drifting pebbles and
grow to planetary sizes 
\citep{Lambrechts_2012,Lambrechts_2014,Guillot_2014}.

Not only the drag on the particles is important, but also the backreaction of the particles on the gas. 
Initially, it was proposed that a secular instability on a settled dust layer
could lead to local particle pileups \citep{Goodman_2000}. 
The pileup would originate from a process resembling plate drag, where the drag
force is assumed to collectively act on a monolithic particle midplane.
This assumption is nevertheless questionable \citep{Youdin_2004} and numerical
studies \citep{Weidenschilling_2006} have not recovered the instability
proposed by \cite{Goodman_2000}.
Nevertheless, this work paved the way for a further investigation on the role
of the backreaction force from gas drag. 
\citet{Youdin_2005} identified a linear instability in the disc midplane.
Their breakthrough result demonstrated that infinitesimal perturbations grow on
an orbital time-scale when the dust-to-gas ratio is around unity or higher.
This instability leads to spontaneous particle clumping, triggering the
gravitational collapse that results in the formation of planetesimals.
In a series of papers \citep{Youdin_2007, Johansen_2007, Johansen_2009,
Johansen_2012} the linear and non-linear evolution of this instability were
numerically investigated in detail. 
These results were independently confirmed and further explored by several
other groups \citep{Bai_2010b, Bai_2010a, Bai_2010c, Miniati_2010, Kowalik_2013}.

Several criteria for the streaming instability to 
achieve particle clumping have been identified:
\begin{itemize}
  \item a disc with slightly supersolar dust-to-gas ratio
    \citep{Johansen_2009,Bai_2010c},
  \item particles of Stokes number $\tau_{\rm f} \sim 0.05$--$0.5$,
    approximately between mm and dm in size
    \citep{Johansen_2007,Bai_2010b,Bai_2010c,Carrera_2015} and 
  \item low radial pressure support in the disc \citep{Bai_2010c}.
\end{itemize}

Further investigations are moving towards a more global understanding of the
effects of the streaming instability, by expanding the simulation domain in the
azimuthal \citep{Kowalik_2013} or vertical direction \citep{Yang_2014}. 
Additionally the streaming instability is placed in a larger context by
incorporating magnetized turbulence \citep{Johansen_2007b}, dust coagulation
models \citep{Drazkowska_2014} or vortex formation \citep{Raettig_2015}.

In this paper we take a step back and study the general process of particle sedimentation in flows with a dust loading comparable to the gas density.
The aim is twofold. Firstly, we hope to gain theoretical insight into particle
sedimentation and more complex drag instabilities, such as the streaming instability and the photoelectric instability \citep{Lyra_2013}.
Secondly, the sedimentation of particles is accessible to laboratory
experiments, thus allowing for a potential experimental confirmation of a particle drag instability.

Of specific interest is the question whether any particle clumping will even
occur at all in a mass-loaded particle rain. 
From previous analytic work on the streaming instability, we do not expect a
linear instability to be present, because of the lack of rotation in a
pure sedimentation problem.
This removes the Coriolis force which is deemed necessary for the streaming
instability to operate \citep{Jacquet_2011, Youdin_2005}.
Nevertheless, in our physical set up (described in Section \ref{sec:model}) 
a non-linear drafting instability is clearly present. The results are described
in Section \ref{sec:numerics}.
The implications of this instability are discussed for particle
sedimentation in protoplanetary discs, chondrule formation, and the envelopes
of giant planets (Section \ref{sec:implications}).
We also place our results in the context of planned laboratory
experiments (Section \ref{sec:lab}). 
We summarise our findings in Section \ref{sec:sum}.

\section{Mass-loaded particle rain}
\label{sec:model}

\subsection{Model equations}

We study the differential motion between particles initially moving with
terminal velocity and stationary gas in hydrostatic balance. 
The dynamics of the gas component is described by
\begin{align}
  \partial_t \rho +\nabla \cdot (\rho \vc u) &=0, 
  \label{eq:SI_gas_cont}\\
  \partial_t \vc u + \vc u \nabla \cdot \vc u &= 
  -g \vc e_z
  -\frac{1}{\rho} \nabla P
  +\frac{1}{t_{\rm f}} \epsilon (\vc v- \vc u)
  + \nu \nabla^2 \vc u,
  \label{eq:SI_gas_mom}
\end{align}
where 
$\rho$ is the gas density, 
$\vc u$ the gas velocity, 
$g$ the gravitational acceleration, 
$P$ the pressure, 
$\epsilon = \rho_{\rm p}/\rho$ is the local dust-to-gas ratio,
$\vc v$ the particle fluid velocity and 
$\nu$ the viscosity.
The drag term from the particles onto the gas depends
on the friction time of the particle \citep[in the Epstein drag regime, ][]{Epstein_1924},
\begin{align}
  t_{\rm f} = \frac{\rho_\bullet R}{\rho v_{\rm th}}\,, 
  \label{eq:fric_t}
\end{align}
where $R,\rho_\bullet$ are the radius and solid density of the particle. The
thermal velocity $v_{\rm th}$ is approximately equal to the local, isothermal gas sound speed
$v_{\rm th} = \sqrt{8/\pi}c_{\rm s}$.

We investigate a regime where we assume an efficient coupling between
particles and gas. 
In this case, the viscous diffusion time for momentum transport between an
average particle pair located a distance $l_{\rm pair}$ apart is shorter than
the friction time of a single particle, 
\begin{align}
t_{\nu,{\rm pair}} = \frac{l_{\rm pair}^2}{\nu} = \frac{n_{\rm p}^{-2/3}}{\nu}
  < t_{\rm f}, 
  \label{eq:t_visc}
\end{align}
with $n_{\rm p}$ the particle number density. 
The particles can then be described by a pressureless fluid
\citep[a formal derivation can be found in][]{Jacquet_2011} with continuity and momentum equation
\begin{align}
  \partial_t \rho_{\rm p} +\nabla \cdot (\rho_{\rm p}\vc v) &=0, 
  \label{eq:SI_part_cont}\\
  \partial_t \vc v + \vc v \nabla \cdot \vc v &= -g \vc e_z
  -\frac{1}{t_{\rm f}} (\vc v- \vc u).
  \label{eq:SI_part_mom}
\end{align}
In the remainder of the paper we will employ `friction units': the friction
time $t_{\rm f}$ as time unit and the friction length $l_{\rm f} = gt_{\rm f}^2$ as length unit.
Velocities can then be expressed in units of terminal velocity $v_{\rm f}=
gt_{\rm f}$. The criterion expressed in Eq.\,(\ref{eq:t_visc}), for
example, reduces to  $ n_{\rm p}'^{-2/3}/\nu'<1$. 
We will preserve the prime notation in the following sections to explicitly denote quantities expressed in friction units.

The use of the friction time as unit of time is possible for the sedimentation
problem, because there are no rotation terms that would necessarily introduce
the additional time scale of the orbital Keplerian frequency, as is for example
the case for the streaming instability \citep{Youdin_2005}.

Expressed in friction units the model equations leave us with only three free
dimensionless parameters:
\begin{itemize}
  \item the viscosity, $\nu' = \nu/(g^2t_{\rm f}^3)$, which is the inverse of
    the Reynolds number ($Re$) in terms of the terminal velocity and the
    friction length,
  \item the sound speed, $c_{\rm s}' = c_{\rm s}/(gt_{\rm f})$, which is the
    inverse of the Mach number ($Ma$) in terms of the terminal velocity, and 
  \item the dust-to-gas ratio, $\epsilon$.
\end{itemize}
The latter is arguably the most important, because we desire to work in the
incompressible limit ($Ma \ll 1$) and we face a lower
bound on the viscosity imposed by the requirement for numerical stability.

A major benefit from this choice of units is that our calculations do not
require us to specify a particle size (Eq.\,\ref{eq:fric_t}). 
Thus our results can be freely scaled to the desired particle size in the
context of protoplanetary discs (Section
\ref{sec:implications}) or a laboratory setting (Section \ref{sec:lab}).

\subsection{Numerical implementation}

In our numerical simulations, performed with the Pencil Code\footnote{The
Pencil Code is open source and can be obtained at \\
\url{http://pencil-code.nordita.org/}. 
A description of the code can be found in
\citet{Brandenburg_2002}, \citet{Brandenburg_2003} and \citet{Youdin_2007}.},
we do not employ the particle fluid
description used for the analytical calculations for our main results, but
instead use a Lagrangian super-particle approach. 
Particles are implemented as super-particles that
represent swarms of physical particles.
This is important, especially in the non-linear regime where it is desirable to
allow particle trajectory crossing and steep density gradients
\citep{Youdin_2007}. 
We have nevertheless used the particle fluid approach also numerically to verify
some of our work in the linear regime.
The assignment of drag forces on the particles and on the gas is further
described in \citet{Youdin_2007} and in \citet{Johansen_2007}. 
Drag is calculated on particles assuming a constant friction time.
We have also made use of the block domain decomposition for particle load
balancing amongst processors \citep{Johansen_2011}.

For the initial condition, we set up a gas column in hydrostatic equilibrium,
taking the drag from the particles on the gas into account.
Particles are typically distributed randomly, with the possibility to add
perturbations (discussed in more detail in Section \ref{sec:nonlin}) and
initiated with with their vertical velocity equal to the terminal velocity.
The gas stratification then takes the form
\begin{align}
  \rho  &= \left( \rho_p +\rho_{\rm b} \right) 
  \exp\left( - \frac{g}{c_s^2} z \right) - \rho_p, 
  \label{eq:hy_stat_par}
\end{align}
see Appendix \ref{ap:linstab} for more details.
This initial condition works well, but we nevertheless find that 
in heavily elongated simulation domains, the stratification of the gas combined
with a uniform particle distribution triggers a vertical gas density
wave, that dissipates over time. There is also a small upwards advection of
gas as the top of the domain gets cleared of particles. 
These effects restricts our simulation domain in practice to approximately
$20$\, $l_{\rm f}$ in the vertical direction (at $c_{\rm s}/v_{\rm f}=10$).

A full list of the performed simulations can be found in Table \ref{tab:runs1}
and Table \ref{tab:runs2}. Below, we describe the nominal numerical set up in
detail.

The boundary condition are set to be periodic in the horizontal direction, for
both the particle and gas component. In the vertical direction particles are
removed from the simulation when crossing the edge of the simulated domain. The
vertical boundary condition on the gas is symmetric (vanishing first
derivative) in all quantities, except for the vertical gas velocity which is
antisymmetric (vanishing value). This boundary condition effectively puts a
solid surface at the bottom of the simulation domain, on which the gas is
supported.

We use an ideal gas equation of state with adiabatic index $\gamma = 5/3$. 
The sound speed is set to be $c_{\rm s} = 10 v_{\rm f}$ (unless mentioned
otherwise) to approach the incompressible limit. 
The Pencil Code is a code optimised for both subsonic and mildly transsonic
flows, but we found a Mach number of $0.1$ sufficient to probe the
incompressible regime of interest.

We found a choice of 16 superparticles per grid cell is sufficiently high to
model a coherent fluid and reduce particle noise (see also Appendix
\ref{ap:np}).
We have standardly used a physical viscosity treatment, but for runs with
extended domain and high mass loading ({\texttt run1.01, run1.02}) we added
sixth order `artificial viscosity' \citep{Haugen_2004}. We employed the minimal
amount necessary to prevent numerical artefacts from developing.
The grid Reynolds number is minimally $32$ times
smaller than the Reynolds number in friction units.
Most simulations were performed in 2 dimensions, but we have verified our
results in 3 dimensions as well ({\texttt run3d.4}, Fig.\,\ref{fig:t_scale}), showing little difference.
Nevertheless, such numerically expensive 3D runs are of interest for further
study.

\begin{table*}
\caption{
Parameters of the numerical simulations. All values are given in the
friction unit system.
Here, $\nu$ is the viscosity, $\epsilon_0$ is the background dust-to-gas ratio,
$c_{\rm s}$ is the sound speed, $A$ and $\lambda$ are the values of the
amplitude and wavelength of the perturbation. The total number of particles is
$N_{\rm par}$ and $N_{\rm cells}$ is the total number of grid cells.}
\label{tab:runs1}
\centering
\begin{tabular}{c c c c c c c c c c}  
\hline\hline  
Name  & $L_x \times L_z$ & Resolution & $\nu$ & $\epsilon_0$ & $c_{\rm s}$ &
Perturb. & $A$ &
$\lambda$ & $N_{\rm par}/N_{\rm cells}$\\
\hline

\texttt{run1} & $1 \times 20$ & $32\times640$  & 1.0e-4 & 1 & 10 & rand & -- &
-- & 16\\
\texttt{run2} & $1 \times 20$ & $64\times1280$  & 1.0e-4 & 1 & 10 & rand & -- &
-- & 16\\
\texttt{run3} & $1 \times 20$ & $128\times2560$  & 1.0e-4 & 1 & 10 & rand & --
& -- & 16\\

\texttt{runRT}& $1 \times 20$ & $32\times640$   & 1.0e-4 & 1 & 10 &
$k_{\rm z}$ & 0.1 & 4 & 16\\
\texttt{runKH}& $2 \times 20$ & $64\times640$   & 1.0e-4 & 1 & 10 & 
$k_{\rm x}$ & 0.1 & 0.5 & 16\\

\texttt{runEGG} & $1 \times 20$ & $32\times640$   & 1.0e-4 & 1 & 10 & eggbox &
0.1 & 1,4 &  16\\
\texttt{runEGG2} & $1 \times 20$ & $64\times1280$  & 1.0e-4 & 1 & 10 & eggbox &
0.1 & 0.5,2 & 16\\

\texttt{runv2} & $1 \times 20$ & $32\times640$  & 1.0e-2 & 1 & 10 & rand & -- &
-- & 16\\
\texttt{runv3} & $1 \times 20$ & $32\times640$  & 1.0e-3 & 1 & 10 & rand & -- &
-- & 16\\
\texttt{runv5} & $1 \times 20$ & $32\times640$  & 1.0e-5 & 1 & 10 & rand & -- &
-- & 16\\
\texttt{runv6} & $1 \times 20$ & $32\times640$  & 1.0e-5 & 1 & 10 & rand & -- &
-- & 16\\

\texttt{run1.e4} & $1 \times 20$ & $32\times640$  & 1.0e-4 & 1 & 10 & rand & -- & -- & 16\\

\texttt{run2.n4} & $1 \times 20$ & $64\times1280$ & 1.0e-4 & 1 & 10 & rand & --
& -- & 4\\
\texttt{run2.n64} & $1 \times 20$ & $64\times1280$  & 1.0e-4 & 1 & 10 & rand &
-- & -- & 64\\

\texttt{run3d.4} & $1 \times 1 \times 20$ & $32\times32\times 640$ & 1.0e-4 & 4
& 10 & rand & -- & -- & 16\\

\hline
\end{tabular}
\end{table*}

\begin{table*}
  \caption{
  Parameters of the numerical simulations extended in the vertical domain
  by $t_{\rm f} = 0.1$.
  Values of variables in friction units (similar to Table \ref{tab:runs1}).
  Here, art. visc. stands for the value of the artificial viscosity parameter.}
\label{tab:runs2}
\centering
\begin{tabular}{c c c c c c c c c c c}  
\hline\hline  
Name  & $L_x \times L_z$ & Resolution & $\nu$ & $\epsilon_0$ & $c_{\rm s}$ & Perturb. & A &
$\lambda$ & $N_{\rm par}/N_{\rm cells}$ & art. visc.\\
\hline
\texttt{run1.01} & $100 \times 2000$ & $32\times640$  & 1.0e-1 & 1 & 100 & rand
& -- & -- & 16 & $10$ \\
\texttt{run2.01} & $100 \times 2000$ & $32\times640$  & 1.0e-1 & 0.25 & 100 &
rand & -- & -- & 16 & $10$ \\
\texttt{run3.01} & $100 \times 2000$ & $32\times640$  & 1.0e-1 & 0.1 & 100 &
rand & -- & -- & 16 & --\\
\texttt{run4.01} & $100 \times 2000$ & $32\times640$  & 1.0e-1 & 0.05 & 100 &
rand & -- & -- & 16 & --\\
\hline
\end{tabular}
\end{table*}

\section{Numerical results showing spontaneous particle concentrations}
\label{sec:numerics}

\begin{figure}[ht!]
  \centering
  \includegraphics{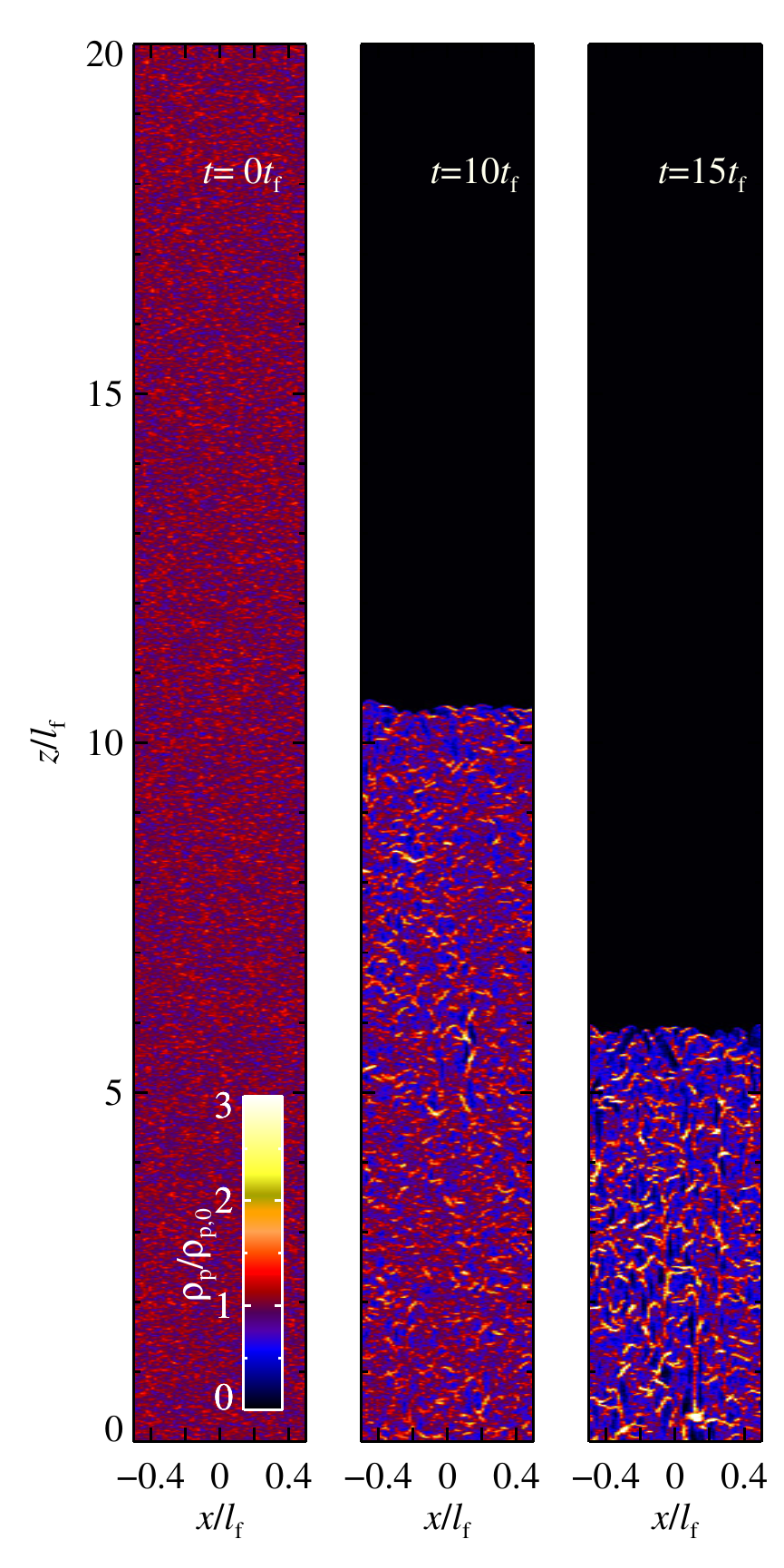}\\
  \caption{
  Development of particle swarms by a drafting instability. 
  Displayed is the evolution of the particle density in the two-dimensional
  simulation \texttt{run2} (see Table \ref{tab:runs1}). 
  The left-most panel illustrates the initial conditions: a stratified gas
  column in the vertical directions with particles sedimenting at terminal
  velocity, placed randomly throughout the simulation domain ($1$\,$l_{\rm f}$
  wide and $20$\,$l_{\rm f}$ high, note that the figure aspect ratio is
  enlarged in the $x$-direction).
  In the following panels (time $t=10,15$\,$t_{\rm f}$), regions marginally
  overdense with particles locally break the stratification equilibrium and
  accelerate downwards, while in particle-poor regions a deceleration from
  terminal velocity occurs. This leads to continued particle pileups through
  drafting, originating from gas being dragged by the particles.
  The particle and fluid components unmix and remain in this state.
  At the end of the simulation, when most particles have sedimented out of the
  simulation domain, the maximal particle density has increased by a factor
  $10$.
  }
  \label{fig:rhop_mxz_panels}
\end{figure}

\subsection{Nonlinear behaviour}
\label{sec:nonlin}

\begin{figure}[t!]
  \centering
  \includegraphics{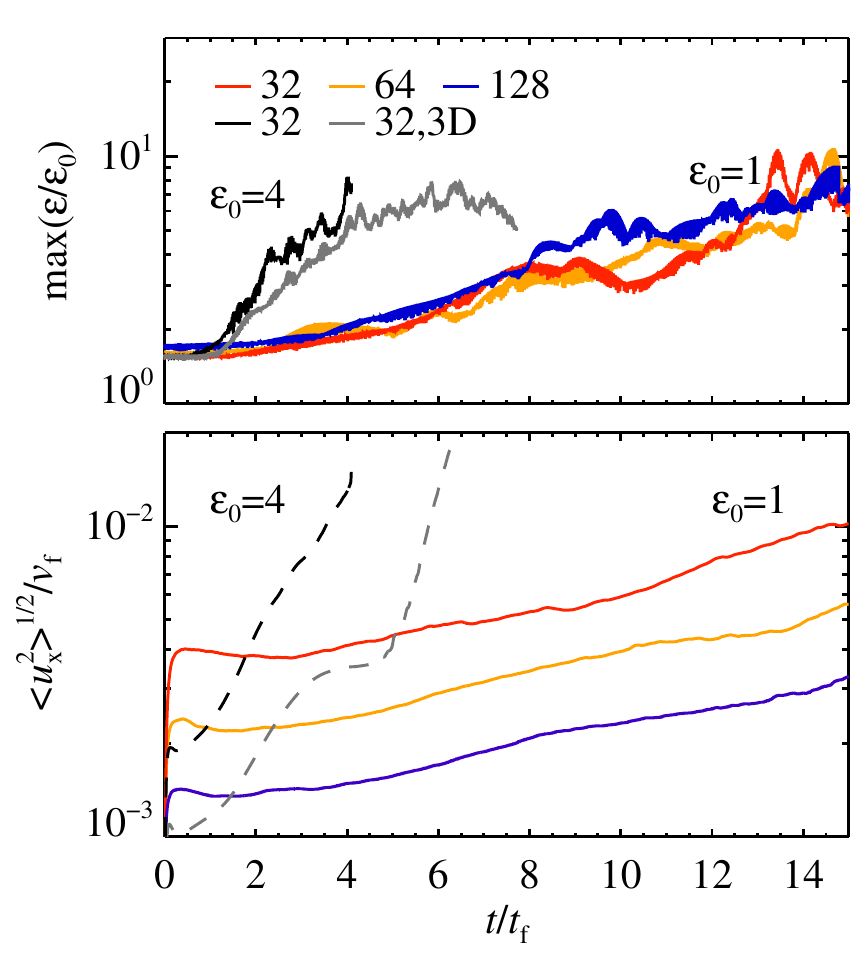}\\ 
  \caption{
  \emph{Top panel:}
  Evolution of the particle density (normalized to the background dust-to-gas
  ratio $\epsilon_0$), for three different resolutions in 2D simulations
   and a 3D simulation (\texttt{run3d.4}, gray).
  \emph{Bottom panel:}
  Evolution of the instability based on the horizontal gas dispersion,
  $\sqrt{<u_x^2>}$, (as an alternative tracer to the maximal particle
  density, which is an intrinsically noisy variable).
  We have used a similar color coding as top panel. 
  Note for the runs with $\epsilon_0$$=$$4$ we have displaced the dashed curves
  for clarity with a factor 10 downwards.
  } 
  \label{fig:t_scale}
\end{figure}

From previous analytic studies of the streaming instability \citep{Youdin_2005,
Jacquet_2011}, it is well known that rotation is an essential ingredient for
the linear phase of particle clumping. The interpretation proposed by
\citet{Jacquet_2011} is that the Coriolis force is necessary to create a
pressure maximum supported by geostrophic balance.
We have repeated the analysis in Appendix\,\ref{ap:linstab} for completeness.
It demonstrates that the sedimentation model without rotation discussed in this
paper is not expected to show a linear instability, under the assumption of
incompressible gas.

Our numerical results nevertheless show that even a minimal disturbance of the
sedimenting particle component with $\epsilon=1$ leads to spontaneous clumping
of material, resulting in the particles to unmix  and sediment out of the fluid. 
Figure \ref{fig:rhop_mxz_panels} illustrates the process. 
Particles located in initially weakly overdense regions sediment faster,
dragging the gas along, resulting in a drafting effect which pulls in more
particles. For clarity, the drafting mentioned here is the result of the
collective motion of a swarm of particles and the resulting gas drag back
reaction, not from individual particle slipstreaming.
At the same time, the opposite occurs in regions less dense than the mean
particle density. 
These effects amplify each other, which results in dense swarms of
particles to form that can undergo secondary instabilities.
For example, one can see in the bottom of the last panel of
Fig.\,\ref{fig:rhop_mxz_panels} a particle cloud resembling a characteristic
Rayleigh-Taylor mushroom, which we will discuss in more detail below.

\subsection{Particle noise perturbation}
\label{ss:pnoise}

We begin by considering the evolution of the sedimenting particles, when they
are placed randomly in the simulation domain. 
This corresponds to a noisy initial condition for the particle distribution,
with maximal relative changes in the particle density on the order of $50$\,\%
for our nominal 2D resolution, for more detail see Appendix \ref{ap:np}.

Figure \ref{fig:t_scale} shows the time evolution of the maximal particle density in the simulated domain (for different background dust-to-gas ratios).
Particle overdensities reach $10$ times the average value, although they are
still growing slowly towards the end of the simulations when the particles fall
out of the box.
Over time, particles sediment out of the simulated domain,
so at late times fewer and fewer particles are traced.
For clarity, we also show the evolution of the horizontal gas velocity
dispersion, which is a less noisy measurement than the particle density. This
illustrates the exponential nature of the instability as well.

Our numerical results stand in contrast to the stable state predicted by
linear stability analysis.
It appears that the main driver for the particle clumping is an imbalanced
stratification. Recall that the particle loading of the sedimenting particles
is taken into account when setting up the equilibrium state
(Eq.\ref{eq:hy_stat_par}). 
This balance can be broken along the $x$-direction by regions with a higher, or
less high, particle density compared to the mean value. 
Apparently, the fluid and particles have no means of finding a global
equilibrium state in the $x$-direction in the response to the particle
fluctuation.
Instead, the particle components breaks into dense swarms.

\begin{figure}[t!]
  \centering
  \includegraphics{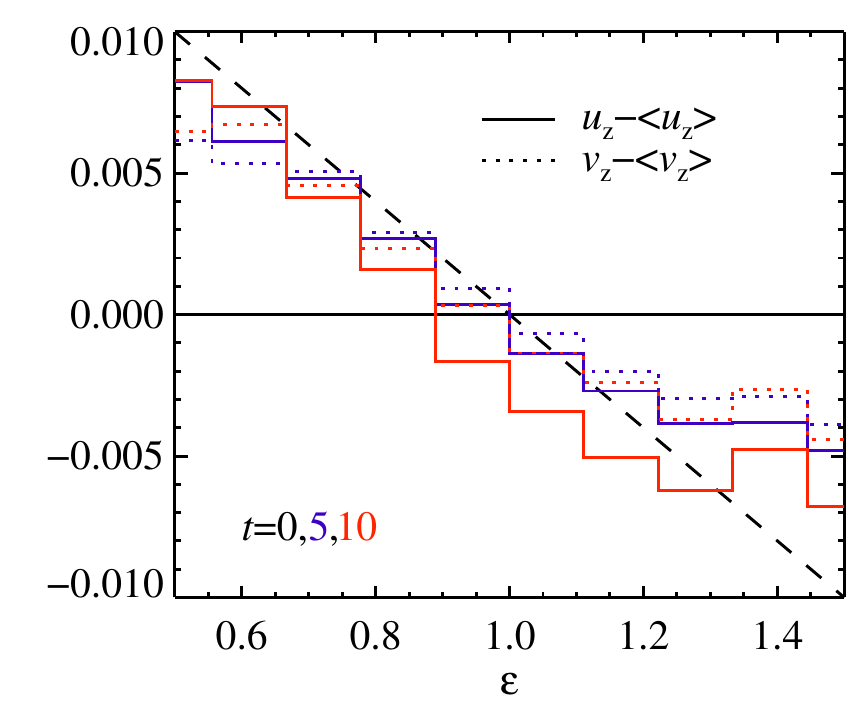}\\
  \caption{
  Correlation between the local dust-to-gas ratio $\epsilon $ and the
  deviations in the velocity of the gas $u_{\rm z}$ (full line) and particles $v_{\rm z}$
  (dotted line).
  To aid interpretation, we subtracted the mean velocity of the
  sedimenting particles $\left<v_{z}\right>$ from the particle velocity and a
  small artificial upwards mean gas velocity $\left<u_{z}\right>$ from the gas
  velocity.
  Displayed are different times $t=0,5,10$\,$t_{\rm f}$, in respectively
  black, blue and red, based on the region between $z$=$0$--$10$\,$l_{\rm f}$ in
  \texttt{run2}. 
  The standard deviations on the binned averages are relatively
  large, $\sigma$$\approx$$0.03$--$0.04$\,$v_{\rm f}$ for the gas and particle
  velocities between $t$=$5$--$10$\,$t_{\rm f}$.
  The black dashed line corresponds to $\alpha=0.02$ in the toy model
  (Sec.\,\ref{sec:toymodel}).
  }
  \label{fig:u_eps_cor}
\end{figure}

This interpretation is supported by the correlation  between
overdense regions and their increased settling speeds shown in
Fig.\,\ref{fig:u_eps_cor}. 
We have binned the surrounding gas and particle velocity in the grid cell
for every particle in the simulation of \texttt{run2}, revealing that on
average gas and particles sediment about $1$\% of faster or slower 
for order unity fluctuations in the dust-to-gas ratio.

From inspection of our numerical results at different resolutions and spatial
scales, we find that the instability tends to originate on the smallest
available scale, near the grid scale in simulations with minimal viscosity
(the dependency on the viscosity is discussed in Sec.\,\ref{sec:Re} in more detail).
This makes it computationally challenging to characterize the instability, as
increased resolutions do not necessarily better resolve the characteristic
scale of interest. Instead, the instability takes place a little faster and
remains quantitatively similar (Fig.\,\ref{fig:t_scale}).

The view of the drafting instability as arising from an imbalanced pressure
stratification suggests that the instability is non-linear, as the particle
fluctuations that drive the horizontal imbalance must be seeded from the
initial condition. This implies that the growth rate of the instability should
decrease with increasing particle number, as randomly placed particles have a
decreased effective density fluctuation with increased particle number. We show
in Appendix \ref{ap:np} that indeed the fastest growth is found when reducing
the particle number to just four per cell. However, we are not able to
completely shut off the instability at larger particle number, indicating that
the instability may operate even in the limit of very high particle number.

In the next sections we investigate how the sedimenting particles
react to different perturbations of the system in order to gain further insight
in the non-linear phase. 
We then propose that a toy model that can capture the dependency of the
instability on the metallicity and Reynolds number (Sec.\,\ref{sec:toymodel} --
\ref{sec:Re}).

\begin{figure}[t!]
  \centering
  \includegraphics{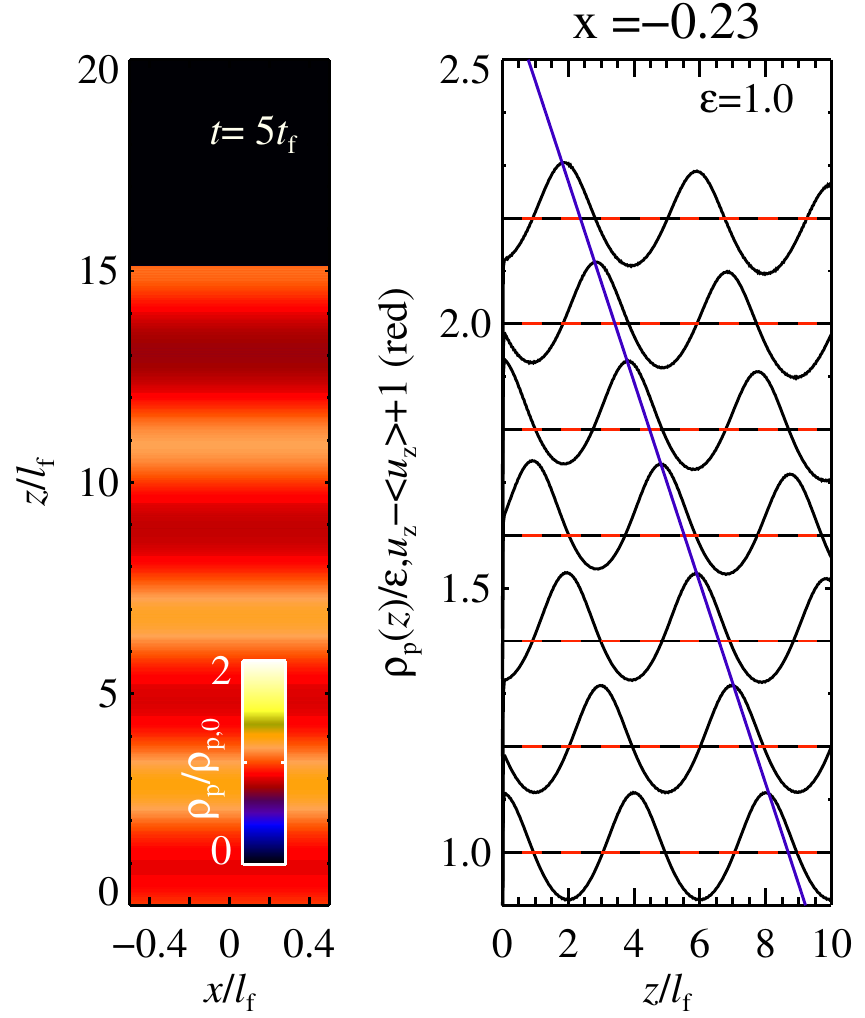}\\
  \caption{ 
  \emph{Left panel}:
  Sedimenting particles with a $k_{\rm z} =\pi/2$-mode in the particle density.
  Represented is a snapshot of \texttt{runRT} at $t=5t_{\rm f}$. No
  instabilities develop, 
  suggesting that the drafting instability is not directly related to the
  Rayleigh-Taylor instabilities that should occur in this set up.
  \emph{Right panel}: 
  Evolution of the sedimenting particle wave at an arbitrary
  placed slice at $x=-0.23$. Different curves represent different times
  ($t=0,1,2,\dots,6$\,$t_{\rm f}$), that
  for clarity are offset by $0.2$.
  The particle density $\rho_{\rm p}(z)$ is given in black, while the red
  dashed line gives the gas velocity $u_z(z)$. 
  Particle ridges do not approach each other, as indicated by the blue line
  that tracks the position of a point advected with velocity $v_{\rm f}$.
  The pressure that is supporting the mass-loaded stratification adapts to
  compensate for small changes in the particle density. This
  allows $u_z(z)$ to remain zero while the particles sediment.
  } 
  \label{fig:ridges_RT}
\end{figure}

\subsection{Wave perturbation}

We now verify numerically that the drafting instability is not related to
either the Rayleigh-Taylor instability or the Kelvin-Helmholtz instability by
perturbing the system with either a purely vertical or purely horizontal a
vertical mode in the particle distribution.

We first present results of vertical wave perturbation of the particle density,
which can be seen in Fig.\,\ref{fig:ridges_RT}.
Within the time that particles sediment out of the domain, no instabilities can
be detected. The right panel of Fig.\,\ref{fig:ridges_RT} shows that the
particles simply sediment at terminal velocity. At the same time, the gas
component does not react to the perturbation. We have verified that this result even stands when feeding additional particle noise to the simulation. We do not see any Rayleigh-Taylor-like instabilities \citep[that occur in a hydrostatic fluid with a dense layer on a lighter one, ][]{Drazin_2004}.
This experiment also demonstrates that the mechanism concentrating particles is
at least two-dimensional. In 1D simulations with only a vertical perturbation
the ridges of increased particle density do not approach each other. 

\begin{figure}[t!]
  \centering
  \includegraphics{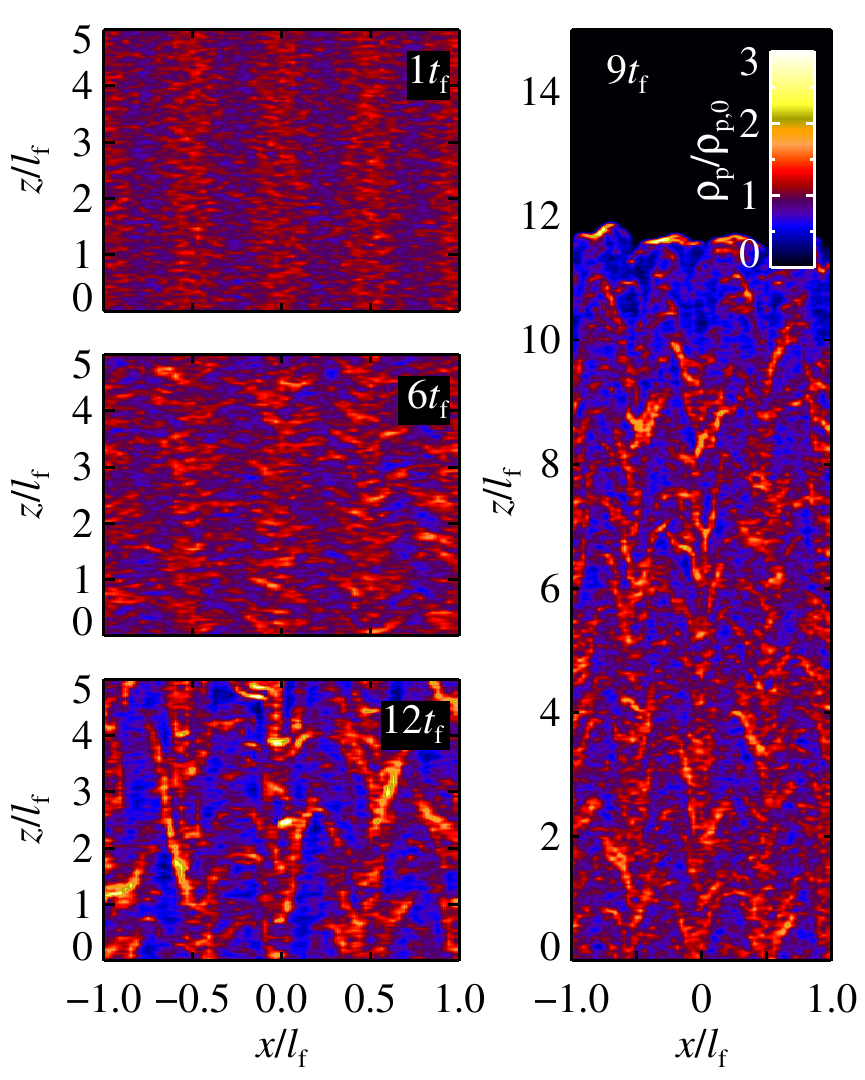}\\
  \caption{
  Sedimenting particles with a $k_{\rm x} =\pi/2$-mode in the particle density.
  We show different details (\emph{left}) from a larger simulated domain
  ({\emph{right}}), at different times ($t=1,6,9,12$\,$t_{\rm f}$), based on
  \texttt{runKH}.
  We added initially noise in the particle velocity perturb the boundary
  between particle rich and poor. 
  We believe that we see a ``parasitic'' Kelvin-Helmholtz instability appear
  over time.
  }
  \label{fig:rhop_u_xz_kh}
\end{figure}

We also experiment with a horizontal wave in the particle
distribution. To perturb the interface between the particle rich and poor
region we displace the particle initially by giving them a random velocity kick
of $v=0.1v_{\rm f}$. 
The resulting evolution is shown in Fig.\,\ref{fig:rhop_u_xz_kh}. The particle dense columns supersediment at an accelerated rate. 
This differential velocity between particle-poor and particle-rich columns
drive an instability reminiscent of the Kelvin-Helmholtz instability
\citep{Drazin_2004}. 
However, in this case the denser fluid that is used in the classical
description of the instability is replaced with a fluid containing an
overdensity in particles.
Such particle-loaded Kelvin-Helmholtz instabilities have been studied before in
the context of molecular clouds \citep{Hendrix_2014}.
A characteristic feature of the Kelvin-Helmholtz instability is the emergence
of v-shaped wings. These can be identified in
the bottom panel of Fig.\,\ref{fig:rhop_u_xz_kh}. 
Similar features are also seen in simulations of the early linear
evolution of the streaming instability in unstratified discs \citep[see for
example Fig.\,2 in][]{Johansen_2007}. Therefore these wings might be a general
feature of particle-gas instabilities.

We expect that this parasitic Kelvin-Helmholtz instability operates in the fully
mixed state of our noise simulations when dense regions start sedimenting out.
It may thus play an important role in the late non-linear evolution. 
However, since we only see this type of behaviour resembling Kelvin-Helmholtz
instabilities in the case of a large perturbation, we do not believe it is the
origin of the instability in the initial noise runs.

\subsection{Eggbox perturbation}

\begin{figure}[t!]
  \centering
  \includegraphics{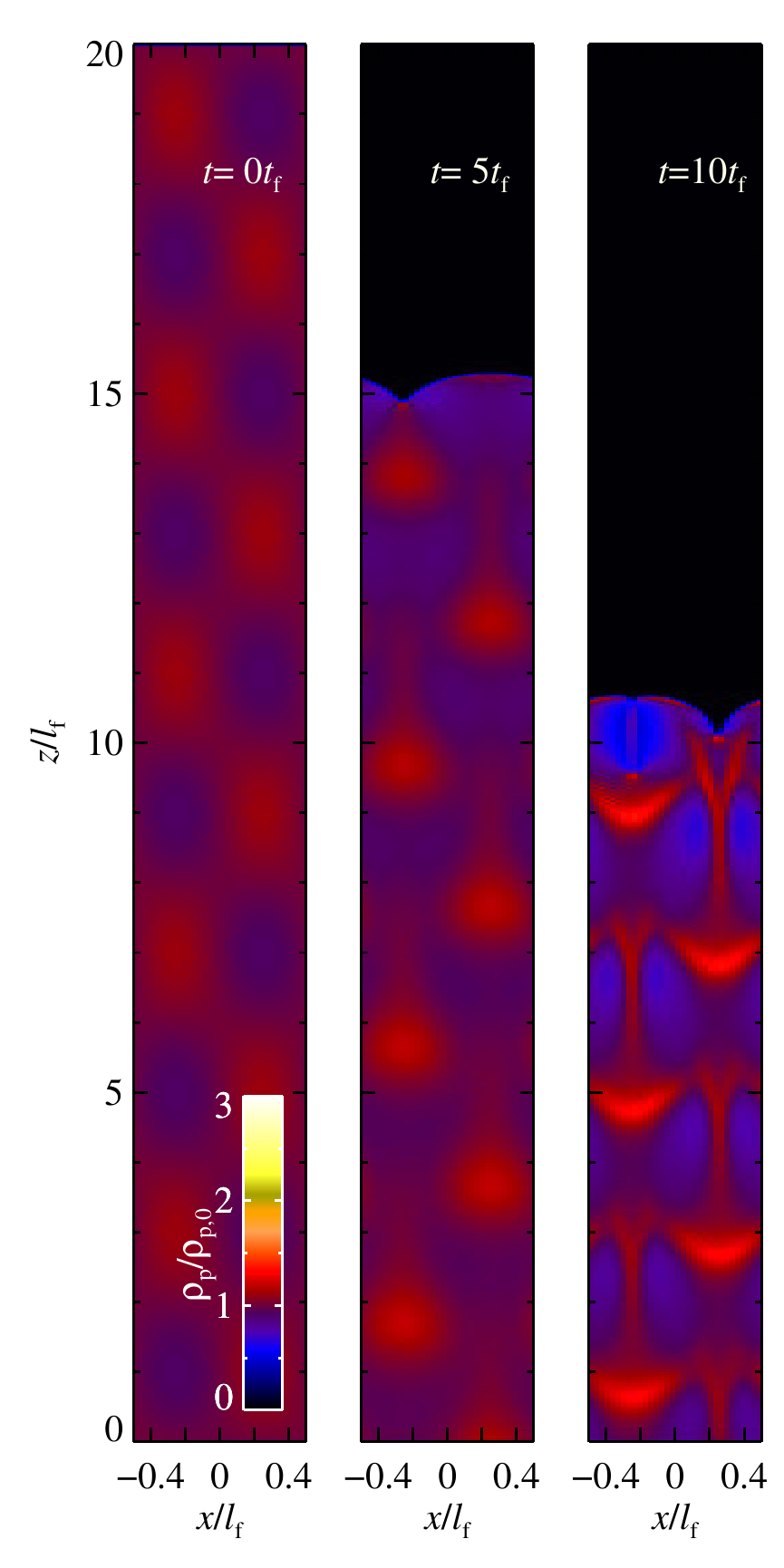}\\
  \caption{
  The evolution of particle droplets, at different times
  ($t=0,5,10$\,$t_{\rm f}$), resulting in the emergence of characteristic
  Rayleigh-Taylor mushroom clouds. 
  The colorbar give the color scale for the shown particle density. 
  Results from \texttt{runEGG}.
  }
  \label{fig:rhop_mxz_panels_eb}
\end{figure}

\begin{figure}[ht!]
  \centering
  \includegraphics{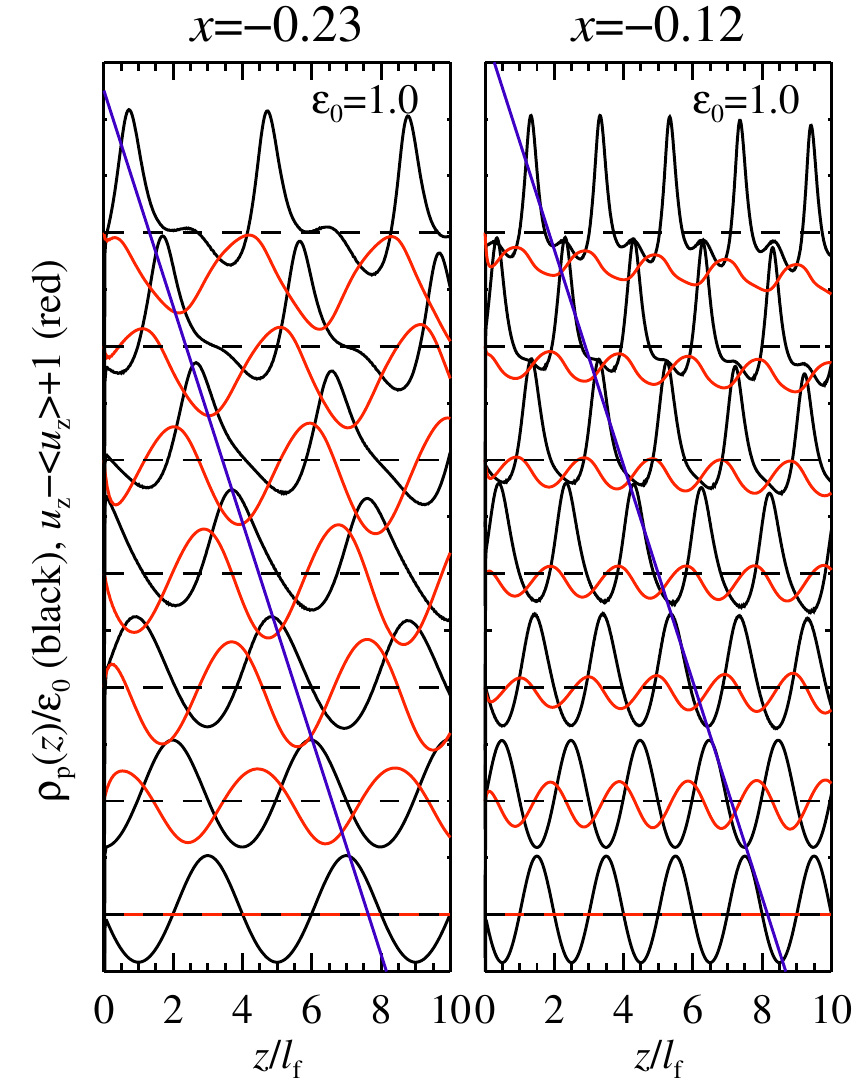}\\
  \caption{
  Slices through the evolution of the particle droplets,  along the
  $z$-direction, similar to Fig.\ref{fig:ridges_RT}.
  The droplets sediment at terminal velocity, and the wave steepening is
  independent of the initial droplet scale. 
  The left panel shows run \texttt{runEGG} with $\lambda_{\rm x}=1,\lambda_{\rm
  z}= 4$ and the right panel \texttt{runEGG2} with $\lambda_{\rm
  x}=0.5,\lambda_{\rm z}=2$.
  }
  \label{fig:ridges_egg}
\end{figure}

Finally, we also perturb the system with an eggbox-like perturbation, in order
to investigate the formation and evolution of particle swarms, which we will
term particle droplets. 
The initial particle density perturbation is of the form
\begin{align}
  \rho_{\rm p}(x,z) = A \sin(k_{\rm x} x) \sin(k_{\rm z} z)\,,
  \label{eq:egg}
\end{align}
with $A$ the amplitude of the two-dimensional perturbation. 
This initial condition can be inspected in Fig.\ref{fig:rhop_mxz_panels_eb}. 
The subsequent panels show the evolution of the inserted particle droplets. Initially
they go through a phase of contraction without altering the amplitude. This can
be seen in further detail in the vertical slices in Fig.\ref{fig:ridges_egg}.
The droplets remain in terminal velocity. 
Intriguingly, the wave steepening is scale-independent, as can be seen the right
panel of Fig.\ref{fig:ridges_egg}, where we have decreased the size of the
droplets by a factor 2. 
Subsequently, the particle transforms in a characteristic mushroom cloud
reminiscent of those seen in the standard Rayleigh-Taylor instability.

\begin{figure}[t!]
  \centering
  \includegraphics{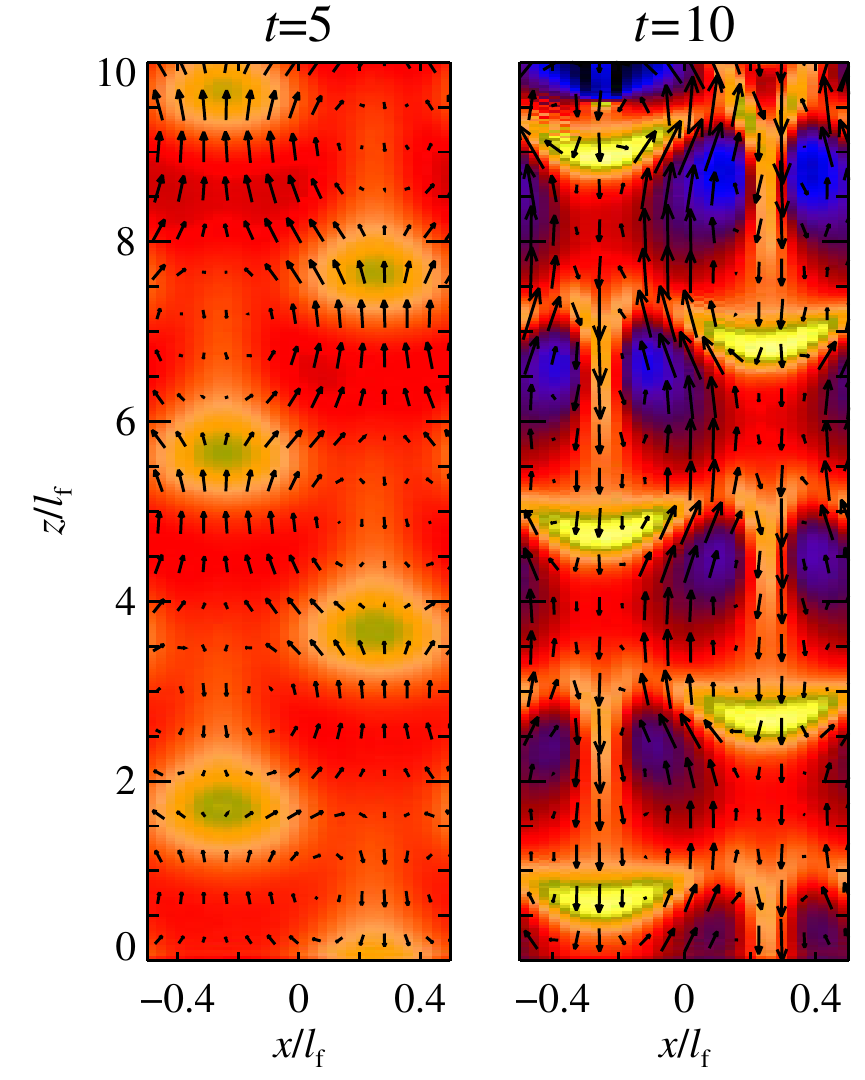}\\
  \caption{
  The particle and gas evolution around particle droplets, at times
  $t=5,10$\,$t_{\rm f}$. Streamlines show the gas velocity, while the particle
  density is color coded as in Fig.\,\ref{fig:rhop_mxz_panels_eb}.
  Results from \texttt{runEGG}.
  }
  \label{fig:rhop_u_xz_egg}
\end{figure}

Figure \ref{fig:rhop_u_xz_egg} illustrates how non-linear drafting
results in particle concentrations. 
Initially, the droplets concentrate the material by collapsing onto themselves.
However, in the subsequent evolution it is clear that the overdense regions
drag gas downwards with their sedimentation flow. 
Simultaneously, gas moves upwards in between the denser areas.  This is
an aspect of the unbalanced stratification that we discussed in
Sec.\,\ref{ss:pnoise}.

\subsection{Toy model of the drafting instability} 
\label{sec:toymodel}

The origin of the instability can be grasped from a simplified
stability analysis, that is based on the observation that regions overdense in
particles sediment faster than regions that are underdense in particles.
Eq.\,(\ref{eq:SI_part_cont}) and Eq.\,(\ref{eq:SI_part_mom}) describe the
behaviour of the particles, which only depends on the gas through the gas drag
term. 
We now assume the gas velocity can be written as 
\begin{align}
  u=\alpha (\epsilon - \epsilon_0) v, 
\end{align}
Here, $\alpha$ is a proportionality constant that can be determined
numerically and which contains the dependence of the instability on
the viscosity and the Mach number. The
quantities $u$ and $v$ are the vertical gas and particle velocity.
Effectively, we use that the linearised gas velocity $\delta u$ can be expressed
proportional to the particle density perturbation $\delta u = \alpha (v_{\rm
f}/\rho_0)\delta \rho_{\rm p}$ (see Appendix \ref{sec:toy} for more details).
Physically, it expresses the observation that gas follows overdense particle
regions, but locally momentum is conserved by the gas becoming buoyant and
moving in the opposite direction in underdense regions.

This assumption allows us to reduce the equations to one dimension, even though
the above approximation implicitly assumes two or three dimensions to be
present to allow gas to move freely and not be trapped as in our stable one
dimensional experiment (Fig.\ref{fig:rhop_u_xz_egg}).
We will also approximate the gas density to be constant.

The equilibrium state corresponds to pure sedimentation with $\epsilon =
\epsilon_0$ and the particles having the terminal velocity $v=-gt_{\rm f}$.
The dispersion relation for Fourier modes of the form $\propto \exp \left(
\omega t -i k x \right)$ becomes
\begin{align}
  \omega' =
  + ik'
  + \frac{1}{2} \left( -1 \pm \sqrt{1 - 4\alpha \epsilon_0 k' i} \right)\,,
  \label{eq:disp_simple}
\end{align}
where $\omega'$ is the growth amplitude, $k' = 2\pi /\lambda'$ the wave number
of wavelength $\lambda'$ in friction units.
The last term of Eq.\,(\ref{eq:disp_simple}) is always positive\footnote{The real part of $\sqrt{1-ix}$ is always larger than 1.} for
$\epsilon_0>0$, resulting in the exponential growth of the instability.  The fastest growing modes are those with the shortest
wavelength.
This result, although derived somewhat differently (see Appendix
\ref{sec:toy}), is identical to that of plate drag
\citep{Goodman_2000,Chiang_2010, Jacquet_2011}.
The real part of the dispersion relation is illustrated in
Fig.\,\ref{fig:growthrate}.

For large $k$, corresponding to short wavelengths ($\lambda \ll
8 \pi \alpha \epsilon_0 t_f^2g$), the growth rate can be approximated by
\begin{align}
  \omega'_{\rm grow} \approx \frac{1}{\sqrt{2}} \sqrt{\alpha \epsilon_0 k'}\,,
  \label{eq:disp_simple_simple}
\end{align}
by series expansion to leading order.
Alternatively, we can express this result no longer in friction units but as 
\begin{align}
  \omega_{\rm grow} \approx \sqrt{\alpha \pi \frac{\epsilon_0 g}{\lambda}}\,.
  \label{eq:disp_simple_simple_real_units}
\end{align}
Interestingly, in this asymptotic limit case the growth rate no longer depends
on the particle size (or more accurately the friction time), but only on
the spatial scale and the dust-to-gas ratio.
On larger scales, the limit expression of the growth rate takes the form
\begin{align}
  \omega'_{\rm grow} = \left(\alpha \epsilon_0 k'\right)^2\,,
  \label{eq:large_scale_limit}
\end{align}
to leading order.
Therefore, the growth rate of the drafting instability rapidly decreases with
increased spatial scale. In this branch the growth rate does depend on the
particle size, 
\begin{align}
\omega_{\rm grow} = (\alpha \epsilon k g)^2 t_{\rm f}^3\,.
\end{align}

\subsection{Dependence on the Reynolds number}
\label{sec:Re}

If the toy model is right, then the growth rate scales as 
$\omega'_{\rm grow} \propto k^{'2}$ on length scales above the characteristic scale
\begin{align}
  \lambda_{\rm knee} = 2^{4/3} \pi\alpha\epsilon_0l_{\rm f},
  \label{eq:knee_scale}
\end{align}
obtained from balancing the fast and slow growth branches
(Eq.\,\ref{eq:disp_simple_simple} and
\ref{eq:large_scale_limit}).

Viscous damping has a similar quadratic dependence on $k'$.
Therefore a viscosity cut-off exists: at $\nu$ larger than $\nu_{\rm crit}$
growth of the instability is terminated.
We find the critical viscosity by equating the large scale growth time scale
(Eq.\,\ref{eq:large_scale_limit})
with the viscous time scale ($\lambda'^{2}/{\nu'}$), 
\begin{align}
  \nu_{\rm crit}' = 2\pi (\alpha \epsilon )^2 \,.
  \label{eq:visc_cut}
\end{align}

\begin{figure}[t!]
  \centering
  \includegraphics{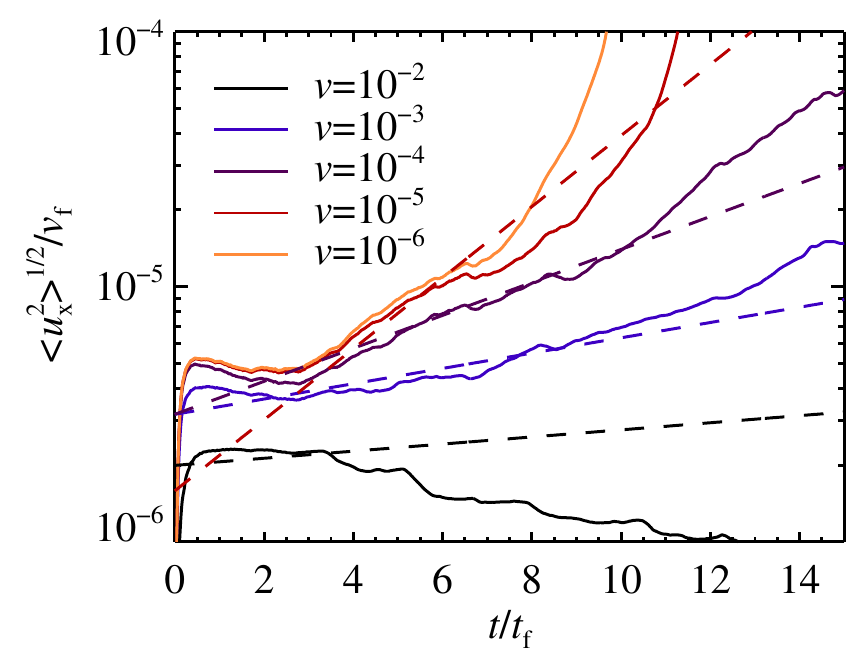}\\
  \caption{
  Viscosity dependency on the growth rate of the drafting instability.
  Dashed lines give the growth time scaling where $\omega \propto \nu^{-1/3}$,
  as in the toy model.
  At high viscosities over $\nu'\approx 10^{-3}$ we no longer identify growth
  in the fluid velocity dispersion.
  Numerical results from \texttt{runv2}, \texttt{runv3},
  \texttt{run2}, \texttt{runv5}, \texttt{runv6}.
  }
  \label{fig:visc_cut}
\end{figure}

The determination of $\alpha$ allows us to scale the growth rates of the
toy model. 
From Fig.\,\ref{fig:visc_cut} we find numerically that above viscosities of
around $\nu' \sim 10^{-3}$, the instability does not show up. 
A critical viscosity of $\nu'_{\rm crit}$$\sim$$10^{-3}$ would correspond
to $\alpha \sim 10^{-2}$. 
Such an estimate is an approximate agreement with the seen correlation between
particle density variations $\epsilon$ and the gas fluid velocity in
Fig.\,\ref{fig:u_eps_cor}.

For viscosities below the viscosity cut-off on the small scale branch, the the
largest growing wavenumber scales as
\begin{align}
 k_{\rm crit}'   = \left( 2 \pi^2 \frac{\alpha \epsilon_0}{\nu'^2} \right)^{1/3}
 \label{eq:k_crit}
\end{align}
by setting the viscous time scale equal to small scale growth rate
(Eq.\,\ref{eq:disp_simple_simple}).
Therefore the growth rate scales with the viscosity as $\omega' \propto \nu'^{-1/3}$.
This indeed agrees with the results shown in Fig.\,\ref{fig:visc_cut}.

\subsection{Dependence on initial dust-to-gas ratio}

\begin{figure*}[ht!]
  \centering
  \includegraphics{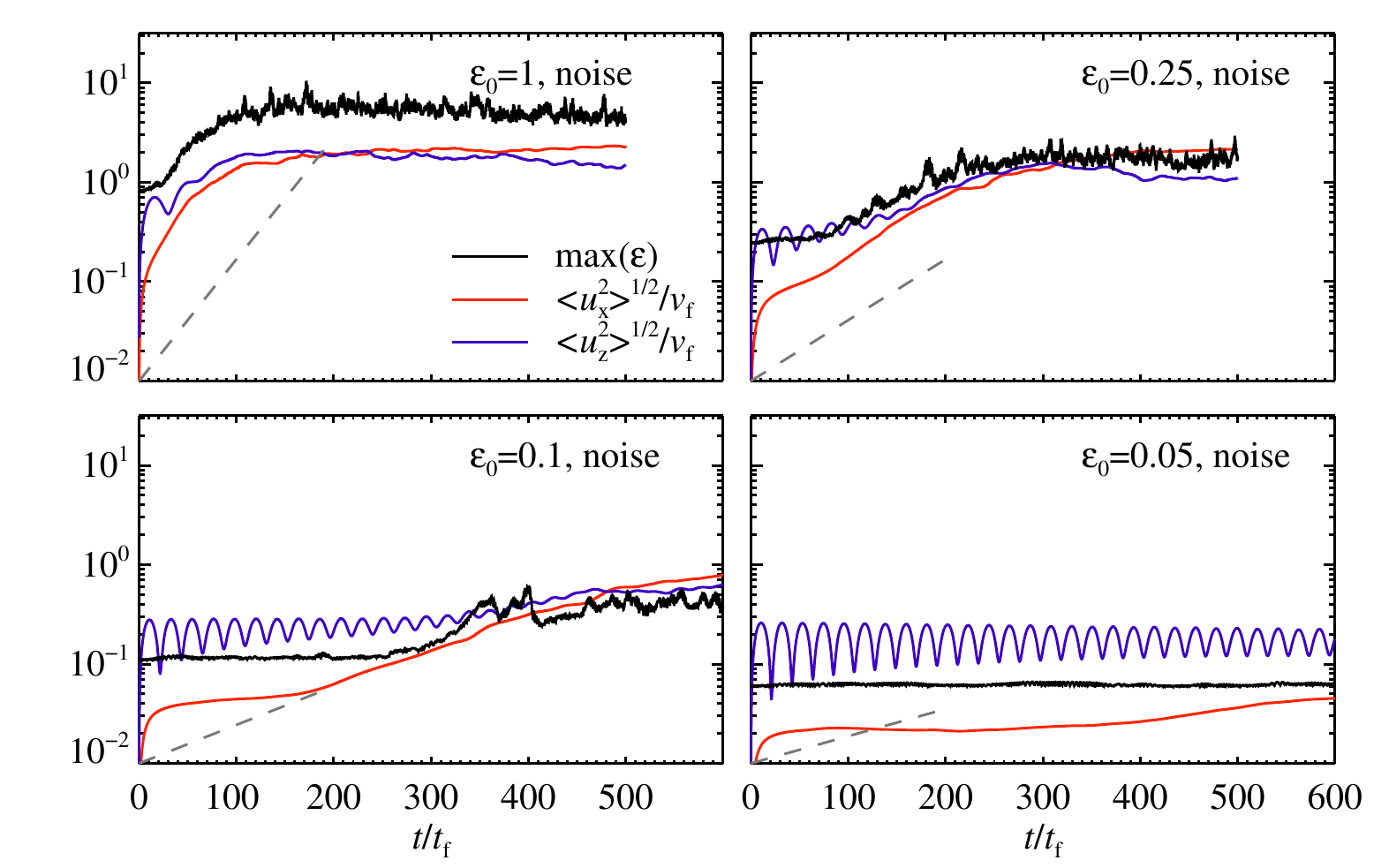}\\
  \caption{
  Long term evolution of the dust-to-gas ratio $\rm{max}(\epsilon)$ (black),
  combined with the horizontal $\sqrt{<u_x^2>}$ (red) and vertical
  $\sqrt{<u_z^2>}$ (blue).  
  The grey dashed lines represent the growth rate scaling with
  $\sqrt{\epsilon_0}$, as found from the toy model
  (Eq.\,\ref{eq:disp_simple_simple}).
  The growth rate decreases with dust-to-gas ratio, and we are unable to measure
  growth rates below $\epsilon_0=0.05$. 
  The small oscillations seen in vertical velocity dispersion for the low
  metallicity runs are the result of vertical waves which are due to a slight
  imbalance in the initial condition in elongated domains.
  The figure is based on the simulations 
  \texttt{run1.01,run2.01,run3.01,run4.01} (in order of decreasing metallicity).
  } 
  \label{fig:t_t01_panels}
\end{figure*}

The main free parameter in our model is the initial dust-to-gas ratio, also
called the metallicity, when setting up the equilibrium stratification
$\epsilon_0$.
Evidently in the limit of negligible dust loading, we do not expect any dust
clumping. 
We therefore study the dependency of the growth rate of the instability on
lower than unity initial dust-to-gas ratio (Fig.\ref{fig:t_t01_panels}). 
To measure slower growth rates (and possibly the saturation of the
instability), we need to extend the vertical domain, which we achieve by
numerically scaling the system (runs \texttt{run1-4.0.1}, see Table
\ref{tab:runs2}).

We find that the instability does not vanish even at a 10 times reduced
metallicity.
The growth rate is slower, and there seems to be a longer dormant phase before
particle concentrations settle in. The reason of this delay for the instability
to kick in is unclear.
Between $\epsilon_0=0.1$ to $\epsilon_0=1$ the growth rates scale approximately
proportionally to $\sqrt{\epsilon_0}$, as expected form the toy model
(Eq.\,\ref{eq:disp_simple_simple_real_units}).

At even lower metallicities, $\epsilon_0=0.05$ we do not recover the
instability. Growth rates become too slow to identify any particle
clumping in the simulation (\texttt{run4.01}).

\section{Enhanced particle concentrations in protoplanetary discs}
\label{sec:implications}

In this section, we rescale our simulation results to the context of the
protoplanetary disc. 
Because the drafting effect seems to prefer higher dust-to-gas ratios than the
percentage level initially expected in a protoplanetary disc, we will consider
particle settling in a disc that has already undergone some grain growth,
resulting in an already partly settled particle midplane
(Sec.\,\ref{ss:rescale} and \ref{ss:toy}).
In mass loaded regions, the swarms created by the drating instability may aid
the formation of chondrules, which we explore in Sec.\,\ref{ss:chodnrules}.
Finally we comment on the relevance of the drafting instability in the
possibly highly dust-enriched envelopes around protoplanets
(Sec.\,\ref{ss:plan}).

\subsection{Rescaling friction units}
\label{ss:rescale}

The friction units employed in our simulations can be readily rescaled to a
protoplanetary disc setting, for a given particle size expressed in Stokes
number
\begin{align}
   \tau_{\rm f} = t_{\rm f}  \Omega_{\rm K},
  \label{eq:Stokes}
\end{align}
where $\Omega_{\rm K}$ is the Keplerian frequency (for the definition of the friction time $t_{\rm f}$, see Eq.\,\ref{eq:fric_t}).
From this definition, a time $t_{\rm f}$ corresponds to a fraction of a Keplerian time scale,
  $t_{\rm f} = \Omega_{\rm K}^{-1} \tau_{\rm f}$ .
Similarly, the friction length $l_{\rm f}$ can be expressed as
\begin{align}
  l_{\rm f} = gt_{\rm f}^2 = z \tau^2_{\rm f},
  \label{eq:length_ppd}
\end{align}
when the gravity is expressed as $\Omega^2 z$. Here, $z$ is the height
above the midplane.
The friction length in the Minimum Mass Solar Nebula
\citep[MMSN, ][]{Hayashi_1981} at the top of a
particle layer of thickness $H_{\rm p}$ can be written as
\begin{align}
  l_{\rm f} \approx 37 \times 10^3
  \left( \frac{H_{\rm p}/H}{0.1}  \right)
  \left( \frac{\tau_{\rm f}}{0.1} \right)^2 
  \left( \frac{r}{5\,{\rm AU}}  \right)^{5/4} \,{\rm km.}
  \label{eq:lf_vert}
\end{align}
This scale strongly depends on the particle size ($\tau_{\rm f}=0.1$ corresponds to a
$2$\,cm particle at an orbital distance of $r \approx 5$\,AU). We have here
assumed that the particle scale height
is a constant fraction of the gas scale height $H$. 

The ratio of the terminal velocity to the sound speed, the Mach number 
\begin{align}
  Ma =  \frac{v_{\rm f}}{c_{\rm s}} = 
  0.01
  \left( \frac{H_{\rm p}/H}{0.1}  \right)
  \left( \frac{\tau_{\rm f}}{0.1} \right),
\end{align}
reveals the incompressible nature of particle sedimentation.

We can ignore the overall rotation of the protoplanetary disc for the small
scales that we consider here.
The Rossby number $Ro = v_{\rm f}/ (\Omega_{\rm K} l_{\rm f})$ takes the form:
$Ro \sim 1 / (\Omega_{\rm K} t_{\rm f})$, when
using friction scales. Therefore, for particles with small Stokes number
$\tau_{\rm f}
= \Omega_{\rm K} t_{\rm f} \ll 1$, rotation is not important, and the rotation-free assumption is valid.

The kinematic molecular viscosity depends on the gas mean free path $\lambda$
in the midplane of the protoplanetary disc as 
\begin{align}
  \nu = \frac{1}{2} c_{\rm s} \lambda\,.
  \label{eq:visc}
\end{align}
The viscosity can then be expressed in friction units as 
\begin{align}
  \frac{\nu}{g^2t_{\rm f}^3} = 1.6\times 10^{-6}
  \left( \frac{H_{\rm p}/H}{0.1}  \right)^{-2}
  \left( \frac{\tau_{\rm f}}{0.1} \right)^{-3}
  \left( \frac{r}{5\,{\rm AU}} \right)^{3/2}\,.
\end{align}
This value does not differ greatly from the nominal value probed in our
numerical work [$\nu/(g^2t_{\rm f}^3)=10^{-4}$, see also the list of
simulations in Table \ref{tab:runs1}]. 
The strong scaling with orbital radius becomes much weaker if one
considers particles of constant radius, as opposed to constant Stokes number.

Finally, we also verify the viscous particle coupling criterion, given by 
Eq.\,(\ref{eq:t_visc}), holds in the MMSN,
\begin{align}
  \frac{t_{\nu {\rm,pair}}}{t_{\rm f}} \approx 
  3.0 \times 10^{-4}
  \left( \frac{\epsilon_0}{0.1} \right)^{-2/3}
  \left( \frac{\tau_{\rm f}}{0.1} \right)
  \left( \frac{r}{5\,{\rm AU}} \right)^{-31/6},
  \label{eq:visc_coup_MMSN}
\end{align}
where $\epsilon_0$ is the approximate mean dust-to-gas ratio in the particle midplane.

\subsection{Applying the toy model}
\label{ss:toy}

With the help of the toy model we can attempt to further constrain where in the
protoplanetary disc the drafting instability can occur.
Because growth rates decrease rapidly at large scales, we only expect the instability to take place on the small scale branch, below the characteristic scale 
$\lambda_{\rm knee}$. From Eq.\,\ref{eq:knee_scale}, we get
\begin{align}
  \lambda_{\rm knee} 
    &\approx 
    290
    \left( \frac{\alpha}{0.01} \right)
    \left( \frac{\epsilon_0}{0.1} \right)
    \left( \frac{H_{\rm p}/H}{0.1} \right)
    \left( \frac{\tau_{\rm f}}{0.1} \right)^2 
    \left( \frac{r}{5\,{\rm AU}} \right)^{5/4} \,{\rm km}. 
  \label{eq:lamkneevert}
\end{align}
The dust-to-gas ratio of $\epsilon_0=0.1$ will be relevant in a  midplane
layer of solids with $H_p/H=0.1$, when the overall metallicity of the
protoplanetary disc is the canonical $Z=0.01$. 
However, we have chosen to keep dust-to-gas ratio $\epsilon_0$ and the height
of the particle layer $H_{\rm p}$ as independent quantities, because we do not
necessarily want to study the conditions of a particle layer settled to
equilibrium.

A lower limit on the scale of the instability is set by 
viscous damping of the instability at scales below the knee,
\begin{align}
  \lambda_{\rm visc} =& 
  \left( \frac{2^2\pi}{\alpha \epsilon_0} \right)^{1/3} 
  \left( \frac{\nu}{g^2t_{\rm f}^3} \right)^{2/3} l_{\rm f}  \nonumber \\
  \approx& 93
  \left( \frac{\alpha}{0.01} \right)^{-1/3}
  \left( \frac{\epsilon_0}{0.1} \right)^{-1/3}
 \left( \frac{H_{\rm p}/H}{0.1} \right)^{-1/3}
  \left( \frac{r}{5\,{\rm AU}} \right)^{9/4}
  \,{\rm km.}
  \label{eq:lamviscvert}
\end{align}
Note this scale, as opposed to the friction length $l_{\rm f}$
(Eq.\,\ref{eq:lf_vert}), does not depend on the particle size.

\begin{figure}[t!]
  \centering
  \includegraphics{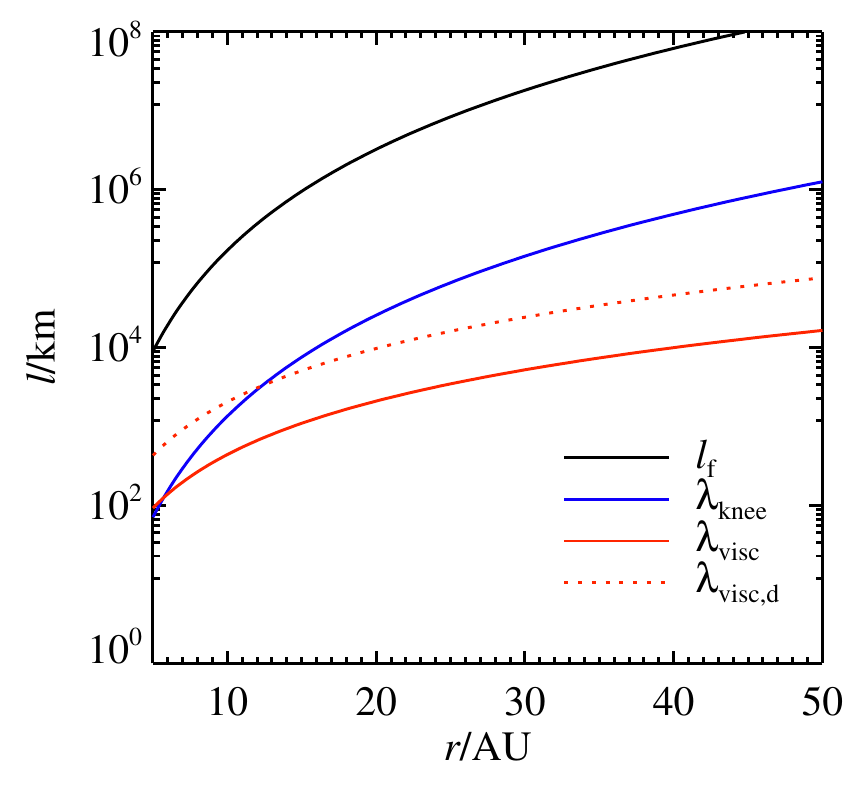}\\
  \caption{
  Relevant length scales for the drafting instability in the Minimum Mass Solar
  Nebula: the friction length ($l_{\rm f}$), the upper scale for fast growth
  ($\lambda_{\rm knee}$), and the scale at which viscosity dominates
  ($\lambda_{\rm visc}$). 
  We consider particles located at a particle scale height above the midplane,
  with $H_{\rm p}/H =0.1$ and the  midplane dust-to-gas ratio is $0.1$. 
  We take the toy model parameter to be $\alpha=0.01$. 
  Particles are assumed to be $1$\,cm in radius, or $1$\,mm in a gas depleted
  disc with 10 times lower gas surface density (in that case the curves remain
  the same, but viscous scale $\lambda_{\rm visc,d}$ is now the red
  dotted line).  
  Likely the instability does not operate in the inner ($< 5$\,AU) of the
  protoplanetary disc, unless particles are large or significant
  pre-concentration occurs.
  }
  \label{fig:timescale_MMSN}
\end{figure}

In Fig.\,\ref{fig:timescale_MMSN} we have illustrated the different relevant
scales presented in Eq.\,(\ref{eq:lf_vert}), (\ref{eq:lamkneevert}) and
(\ref{eq:lamviscvert}), as function of orbital radius. 
The instability would operate on a scale of the order of $10^4$\,km in the
outer parts of the protoplanetary disc for particles of cm in size, assuming
a MMSN model.
Note that Fig.\,\ref{fig:timescale_MMSN} shows the scaling for an assumed
constant particle size, as opposed to
Eq.\,(\ref{eq:lf_vert}--\ref{eq:lamviscvert}) that assume constant Stokes
number.

\subsection{Chondrules}
\label{ss:chodnrules}

Chondrules are mm-sized inclusions found in primitive meteorites originating
from the asteroid belt. 
It is generally accepted that a chondrule is the product of a flash heating
event. 
The exact nature of chondrule precursors is unknown. 
However the heating events likely occurred in particle swarms at
least $100$ to $1000$\,km wide, with a local number density of about
$\sim$10\,m$^{-3}$. In this way the loss of light isotopes (isotopic
fractionation) is prevented by exchanging vapour from chondrule to chondrule
\citep{Cuzzi_2006}.
This scenario requires local chondrule densities more than 100
times above a dust-to-gas ratio of unity.
Even higher concentrations might be necessary to explain the retention of
sodium \citep{Alexander_2008}.

Such high chondrule densities are surprising, since small particles are
hard to concentrate to the midplane. 
Even in the absence of other forms of turbulence, particles sediment to a
midplane with dust-to-gas ratio not higher than approximately unity, because of
the stirring caused by the streaming instability \citep{Bai_2010b}.
However, the isotopic constraints on the need to concentrate chondrules weaken
if the gas at the chondrule formation sites had a non-solar composition.
The atmospheres around planetary embryos have been proposed to be such locations
\citep{Morris_2012}. 
Nevertheless, in this scenario, pre-clumping of solids by a factor of at least
$10$ over midplane densities remains necessary and the shock waves invoked to
melt chondrules lead in fact to destructive collisions \citep{Jacquet_2014}.

Small particles are difficult to concentrate in the inner protoplanetary
disc, because of the strong sensitivity of the preferential scale of the
instability on particle size (Eq.\,\ref{eq:lf_vert} and
Eq.\,\ref{eq:lamkneevert}), as can be seen in Fig.\,\ref{fig:timescale_MMSN}.
Nevertheless the connection to chondrule formation is tantalizing, especially
because if clumping conditions are met, the drafting effect only weakly depends
on particle size and efficiently clusters particles down to very small sizes
(Eq.\,\ref{eq:disp_simple_simple_real_units}). This is different from, for
example, the streaming instability that has a preferred particle size, somewhat
above that of chondrules for nominal metallicities \citep{Carrera_2015}.

The drafting instability could operate on such small scales, if some form of
pre-concentration of solids would occur. Possibly such enhanced particle
densities could occur near the Kolmogorov scale of the disc turbulence
\citep{Cuzzi_2001}. Alternatively, near sublimation lines particle
concentrations can dramatically peak \citep{Ros_2013}. An increase in the
dust-to-gas ratio can also occur by accretion of gas onto the
star, which depletes the disc relative to the MMSN \citep{Bitsch_2015}. 
Alternatively, growth rates could be increased if the unknown chondrule
precursors are much larger than the chondrules they are turned into after the
heating event. 
Even so, it remains to be seen if drafting instabilities can push particle
concentrations to the desired high levels, even in such favourable instances.

\subsection{Planetary atmospheres}
\label{ss:plan}

The drafting instability might be important in the atmospheres of giant
planets. The opacity in the outer envelope, which regulates the transport of
heat, comes from the dust component. 
Under standardly assumed opacities, it is difficult to cool the
envelope and trigger runaway gas accretion \citep{Ikoma_2000,Piso_2014}.
However, clumping of solids and the growth of the accreted dust could
significantly reduce the opacity in the upper atmosphere. 
The friction length for particles sedimenting in a planetary atmosphere is
given by
\begin{align}
  l_{\rm f,plan} &= \frac{GM}{r_{\rm B}^2} t_{\rm f}^2  \nonumber \\
  &\approx 7.2 \times 10^3
  \left( \frac{R}{1\,{\rm mm}} \right)^{2}
  \left( \frac{M}{5\,{\rm M}_{\rm E}} \right)^{-1}
  \left( \frac{r}{5\,{\rm AU}} \right)^{5}
  \,{\rm km}\,,
  \label{eq:lgplan}
\end{align}
where $r_{\rm B} = GM/c_{\rm s}^2$ is the thermal Bondi radius
of a planet with mass $M$, corresponding to the outer edge
of the atmosphere. We have here considered particles on top of the envelope,
but deeper in the planet the friction time shrinks due to the increase in
density. 
Applying the toy model, we estimate the knee scale in the upper envelope at 
\begin{align}
  \lambda_{\rm knee,plan} \approx 
  570 
  \left( \frac{\alpha}{0.01} \right) 
  \left( \frac{\epsilon_0}{1} \right)
  \left( \frac{R}{1\,{\rm mm}} \right)^{2}
  \left( \frac{M}{5\,{\rm M}_{\rm E}} \right)^{-1}
  \left( \frac{r}{5\,{\rm AU}} \right)^{5}
  \,{\rm km}\,,
  \label{eq:lkneeplan}
\end{align}
which is above the damping viscosity scale at
\begin{align}
  \lambda_{\rm visc,plan} \approx 13
  \left( \frac{\alpha}{0.01} \right)^{-1/3}
  \left( \frac{\epsilon_0}{1} \right)^{-1/3}
  \left( \frac{M}{5\,{\rm M}_{\rm E}} \right)^{1/3}
  \left( \frac{r}{5\,{\rm AU}} \right)^{2}
  \,{\rm km.}
\label{eq:viscplan}
\end{align}
The formation of ice giants and super-Earths might be paired with significant
amounts of dust in their low-mass gaseous envelopes \citep{Lee_2014}.
These scaling relations argue that order-of-unity mass loading of
atmospheres will lead to clumping and the breakup of the dust component, 
providing an upper limit on the dust opacity.

\section{Future outlook}

\subsection{Laboratory Experiments}
\label{sec:lab}

This study supports ongoing work to investigate drag instabilities in the
laboratory. 
A full description of the apparatus constructed at the Max Planck Institute
for Dynamics and Self-Organisation and first results will be presented in an
accompanying paper (Capelo et al, 2016, \emph{in prep}). Here, we restrict
ourselves to highlight some aspects relevant to the understanding of the numerical results presented in this paper.

The experimental apparatus consists of a cylindrical vessel, housing a gas
stream operating at pressures in the range of $0.5-10^{3}$\,millibar. 
The axial component of the cylindrical flow is parallel with the direction of
Earth's gravity, similar to the sedimentation configuration in the simulations
presented here.
The upwards steady-state flow will be seeded with weakly inertial
particles, with typical sizes of $20$-$90$\,$\mu$m.
The range of operational pressures and temperatures, listed in Table
\ref{tab_apparatus},  will then allow us to span both the Stokes and Epstein
drag regimes.

The particle entrainment happens upstream in the fluid flow. There the system
is in a brief transient state. The solids are transported and
mixed with the gas by the time the flow reaches the steady-state conditions in
which the measurements are to be made. 
This is done to make a fair comparison with the nearly homogeneously mixed
initial conditions of the two-fluid dust/gas models. 

Table \ref{tab_apparatus} summarises the parameter region in which the
experiment operates, including gas state variables, Mach and Reynolds
numbers. 
The flow conditions are incompressible and laminar. 
The experiment will be the first in its kind probing the Epstein drag regime in
a fluid with equal mass loading of gas and particles. 

The experiment described here is somewhat analogous to particle suspension
experiments in Newtonian fluids with low particle Reynolds number \citep{Guazzelli_2011}. 
However, in these studies volume fractions, $\phi = n_{\rm p}R^3$ (with
$n_{\rm p}$ and $R$ the particle number density and radius), are no lower
than $\phi \approx 0.01\%$. Our experiment operates at $\phi\approx
10^{-4} \%$, when the dust-to-gas ratio is unity, for solid spherical particles
with densities ranging from that of vitreous carbon
($\rho_\bullet$=1.4\,g\,cm$^{-3}$) to steel ($\rho_\bullet$=8\,g\,cm$^{-3}$).
The low particle Reynolds number, $Re_{\rm p}$, in such suspension experiments
comes from the use of a fluid with high dynamic viscosity. The particles are
very buoyant and slowly creep through a thick liquid. Here, on the other hand,
the low values of $Re_{\rm p}$ come from the fact that the kinematic viscosity
becomes high when the gas density is low. 
It is encouraging that such experiments, even if in a regime different from the one studied here, show interesting particle dynamics \citep{Batchelor_1972}, such as particle Rayleigh-Taylor mushrooms and drafting particle trains \citep{
Pignatel_2009, Matas_2004}.

\begin{table}
\caption{Parameters of the laboratory experiment.
The range of pressure values correspond to different settings used to seed
particles of various sizes and densities in the flow. 
The range in temperature values corresponds to the cooling that occurs as the
gas expands to reach steady low-pressure conditions. 
The Reynolds numbers are calculated using the definition, $Re = \rho v L/\mu$,
where $\rho$ is the density of the gas, $v$ the characteristic velocity, $L$
the characteristic length scale, and $\mu_{\rm
air}=1.8\times10^{-5}$\,kg\,m$^{-1}$\,s$^{-1}$ is the dynamic viscosity of air
at room temperature. 
For the particle Reynolds number, we take the characteristic velocity and size
to be the terminal velocity and the particle diameters, respectively. 
The density of the gas is estimated using the measured values of temperature
and pressure, assuming a mean molar mass of air $M_{\rm air}=0.02891$\,kg\,mol$^{-1}$
and molar gas constant $R=8.314$\,m$^{3}$\,Pa\,K$^{-1}$\,mol$^{-1}$. 
The global Reynolds number comes from the mean flow velocity and the tube
diameter. 
Similarly, the Mach number is the ratio of the mean flow velocity to the sound
speed at the measured temperature, again assuming the same values of $R$ and
$M_{\rm air}$.  }
\label{tab_apparatus}
\centering
\begin{tabular}{ l l c l c }  
\hline\hline                  
Property & Value &  \\
\hline

Working gas & air &     \\
Tube height & 1.6 m&  \\
Tube diameter & 9 cm&\\
Friction time & 0.05-0.08 s&\\
Friction length  & $\approx$ 3-7 cm&\\
Pressure range &10-8000 Pa&\\
Temperature & 16-22$^{\circ}$C&\\
Estimated mean flow speed & 1.2 m s$^{-1}$&\\
Global Reynolds number & 0.6 - 6 &\\
Particle Reynolds number & 0.009-0.08 & \\
Mach number & 0.003 & \\
Solid-to-gas ratio & $0.1$--$10$ & \\

\hline
\end{tabular}
\end{table}

Time-resolved data on the particle trajectories will be obtained from
high-resolution cameras and 3-dimensional Lagrangian particle tracking \citep{Xu_2008b,Ouellette_2006}. 
This is a common technique to study both tracer and intertial
particles in fluids. The typical measured and derived quantities are the
probability density distributions of the particle velocities and accelerations,
their statistical moments, and correlation and structure functions.

The obtained data will provide an interesting comparison to the results shown in
this paper. 
The parameter regime is sufficiently similar to the numerical experiments
that we expect the drafting instability to manifest itself. 
Particle tracking would not only allow the detection of particle swarms, but
also the individual particle dynamics. 
For instance, Fig.\,\ref{fig:t_scale} demonstrates that the growing maximum in
particle velocity dispersion traces the increase in maximum particle density.
Such statistical measurements of the particles will allow qualitative comparison between the numerical work and the experiments.

\subsection{Numerical work}

We have here presented several numeric experiments to demonstrate a drafting
instability. Future work will refine the estimates made in this paper.

For example, currently the numerical set up is limited to studying sedimentation
on rather short timescales, set by the length of the simulation domain. 
This could be avoided in future work by implementing a form of periodic boundary
conditions in the vertical direction, which would recycle particles.

To aid the interpretation of the experimental results, it will be necessary to
specifically reproduce the parameter regime in which the apparatus operates.
Additionally, refined boundary conditions will be needed to approximate the
experimental set-up.  Such work is under progress, but evidently awaits the
first experiments.

We have also argued that the drafting instability could be of relevance in a
protoplanetary disc. To study this connection in more detail, it will be
necessary to simulate numerically expensive larger domains encompassing the disc
midplane. 
Additionally, the connection between the drafting and streaming
instability could be studied in more detail. 
Ultimately, the results should be placed in the context of other sources of disc
turbulence, such as the magnetorotational instability operating in
sufficiently ionised regions or in the penetration of vertical shear instability
to the midplane \citep{Turner_2014}.
Additionally, the growth of particles through coagulation or condensation will need to be taken into account in a self-consistent matter.
Future work is needed to understand the possibly constructive interplay of
these mechanisms.

\section{Summary}
\label{sec:sum}

In this paper we have demonstrated the presence of a drafting instability when
particles sediment through a fluid in hydrostatic balance. 
On time scales of tens of friction times particle unmix out of a homogeneous
mixture and particle concentrations increase by a
factor $10$.

The presence of such an instability was not expected because it evades detection
in an analytic linear stability analysis.
However, our numerical results demonstrate that the system is non-linearly
unstable. 
The exact nature of the instability is difficult to determine. We interpret the
instability to be the result of an imbalance in the stratification locally
disturbing the hydrostatic balance. We support this hypothesis with a simple
toy model that captures some of the main characteristics of the instability.
Growth is fastest at the smallest available scales, increases with the square
root of the dust-to-gas ratio and a critical scale is identified at which
viscosity overwhelms the instability. 

By expressing our numerical results in a system of friction units, we can
exploit our results by scaling them to either upcoming laboratory experiments or
the protoplanetary disc. We argue that an experiment can probe a similar regime
with dust-to-gas ratio around unity that is of interest here.
In protoplanetary discs the drafting instability may take place in particle-rich
layers above the midplane in the outer regions of the disc, on scales smaller
than previously studied.
In these regions where the conditions for the drafting
instability are met, we have shown that sedimenting particles spontaneously
form dense clumps. Future work will be needed to investigate to what degree
this clumping affects coagulation rates and whether the drafting instability
can create the dense environment necessary for chondrule flash heating.

\begin{acknowledgements}
  We would like to thank Haitao Xu, Wlad Lyra, Chao-Chin Yang, Karsten
  Dittrich, Hubert Klahr, Neal Turner, Mario Flock and Andrew Youdin for
  valuable comments.
  The paper benefited greatly from insightful comments from two referees, who
  helped clarify the stability analysis and the interpretation of the
  laboratory constraints on chondrule formation densities.
  M.L. acknowledges funding from the Knut and Alice Wallenberg Foundation.
  AJ acknowledges financial support from the European Research Council (ERC
  Starting Grant 278675- PEBBLE2PLANET), the Swedish Research Council (grant
  2010-3710) and the Knut and Alice Wallenberg Foundation (grant "Bottlenecks
  for particle growth in turbulent aerosols" Dnr. KAW 2014.0048)".
  H.\,C., J\,.B. and E\,.B.\,acknowledge support by the Deutsche
  Forschungsgemeinschaft under the grant INST 186/959-1 as part of the CRC 963
  ``Astrophysical Flow Instabilities and Turbulence''.
\end{acknowledgements}

\appendix 

\section{Linear stability analysis}
\label{ap:linstab}

We briefly rederive the stability analysis for a particle gas-mixture in a
non-rotating flow \citep{Youdin_2005, Jacquet_2011}.
We demonstrate the result in two spatial dimensions, but our conclusions remain
valid when generalized to three dimensions.

\subsection{Governing equations}
We assume the gas to be incompressible, in line with \citet{Youdin_2005},
\begin{align}
  \frac{\partial u_x }{\partial x} + \frac{\partial u_z}{\partial z} = 0,
  \label{eq:cont_g_2Di_inco}
\end{align}
and use the standard momentum equations
\begin{align}
  \frac{\partial u_x}{\partial t} 
  + u_x \frac{\partial u_x}{\partial x} 
  + u_z \frac{\partial u_x}{\partial z} 
  &= - \frac{1}{\rho} \frac{\partial P}{\partial x}
  + \frac{1}{t_{\rm f}} \frac{\rho_{\rm p}}{\rho} \left( v_x-u_x \right)\,,\\
  \frac{\partial u_z}{\partial t} 
  + u_x \frac{\partial u_z}{\partial x} 
  + u_z \frac{\partial u_z}{\partial z} 
  &= -g -\frac{1}{\rho} \frac{\partial P}{\partial z} +
  \frac{1}{t_{\rm f}}      \frac{\rho_{\rm p}}{\rho} \left( v_z-u_z \right) .
  \label{eq:mom_g_2D}
\end{align}

Similarly, for the particle fluid we make use of the continuity equation, 
\begin{align} 
  \frac{\partial \rho_{\rm p}}{\partial t} +
  \frac{\partial}{\partial x}\left( \rho_{\rm p} v_x \right) +
  \frac{\partial}{\partial z}\left( \rho_{\rm p} v_z \right) = 0,
  \label{eq:cont_p_2D} 
\end{align}
and the set of momentum equations
\begin{align}
  \frac{\partial v_x}{\partial t} 
  + v_x \frac{\partial v_x}{\partial x} 
  + v_z \frac{\partial v_x}{\partial z} 
  &= - \frac{1}{t_{\rm f}} \left( v_x-u_x \right)\,, \\
  \frac{\partial v_z}{\partial t} 
  + v_x \frac{\partial v_z}{\partial x} 
  + v_z \frac{\partial v_z}{\partial z} 
  &= -g - \frac{1}{t_{\rm f}} \left( v_z-u_z \right).
  \label{eq:mom_p_2D}
\end{align}
This completes the model with 6 parameters ($\rho_{\rm p}, \rho, v_x, v_z,
u_x, u_z$), and as many equations.

\subsection{Equilibrium solution}

In equilibrium we have no vertical motion $(u_x = 0,v_x =0)$. Additionally, we
assume the gas to be in the rest frame, $u_z= 0$, and particles to be
initially uniformly spread ($\rho_{\rm p} = \rho_{\rm p,0}$ constant). 
This leaves the particle
continuity and $z$-momentum equations as the non-trivial equations determining
$v_z$ and $\rho, \rho_{\rm p}$,
\begin{align}
    \frac{\partial}{\partial z}\left( \rho_{\rm p} v_z \right) 
    &= 0\,, \\
    v_z \frac{\partial v_z}{\partial z} 
    &= -g - \frac{1}{t_{\rm f}} v_z\,,  \\ 
    0 
    &= -g -\frac{c_{\rm s}^2}{\rho} \frac{\partial \rho}{\partial z} +
    \frac{1}{t_{\rm f}}  \frac{\rho_{\rm p}}{\rho} v_z.
  \label{eq:eq_test}
\end{align}
In the last equation we have assumed an isothermal gas,
$P =\rho c_{\rm s}^2$. 
The equilibrium solution then takes the form
\begin{align}
  v_z   &= -gt_{\rm f}\,, \\
  \rho  &= \left( \rho_{\rm p} +\rho_{\rm b} \right) 
  \exp\left( - \frac{g}{c_{\rm s}^2} z \right) - \rho_{\rm p}, 
  \label{eq:hydrostat}
\end{align}
where $\rho_{\rm b}$ is the gas density at the
$z=0$ boundary. 

\subsection{Dispersion relation}

We now consider a first order perturbation of this equilibrium state. For the
gas we find
\begin{align}
  \frac{\partial u_x' }{\partial x} + \frac{\partial u_z'}{\partial z} 
  &= 0 \,,
  \label{eq:cont_g_2D_O1} \\
  \frac{\partial u_x'}{\partial t}
  &= 
  - \frac{c_{\rm s}^2}{\rho_0} \frac{\partial  \rho'}{\partial x}
  + \frac{1}{t_{\rm f}} \frac{\rho_{\rm p,0}}{\rho_0} \left( v_x' -u_x' \right)
  \,,
  \label{eq:mom_g_2D_O1} \\
  \frac{\partial u_z'}{\partial t }
  &= - \frac{c_{\rm s}^2}{\rho_0} \frac{\partial \rho'}{\partial z} 
     -\frac{g}{\rho_0} \rho'
     -\frac{g}{\rho_0} \rho_p'
     + \frac{1}{t_{\rm f}}  \frac{\rho_{\rm p,0}}{\rho_0}
     \left( v_z' -u_z' \right).
  \label{eq:mon_g_2D,z}
\end{align}
For the particles we get
\begin{align}
  \frac{\partial \rho_{\rm p}'}{\partial t} +
  \rho_{\rm p,0} \frac{\partial v_x'}{\partial x} 
  +\rho_{\rm p,0} \frac{\partial v_z'}{\partial z}
  + v_{z,0} \frac{\partial \rho_{\rm p}'}{\partial z}
  &= 0\,,
  \label{eq:cont_p_2D_O1} \\
  \frac{\partial v_x'}{\partial t} 
  + v_{z,0} \frac{\partial v_x'}{\partial z} 
  &= - \frac{1}{t_{\rm f}} \left( v_x'-u_x' \right) \,,\\
  \frac{\partial v_z'}{\partial t} 
  + v_{z,0} \frac{\partial v_z'}{\partial z} 
  &=  - \frac{1}{t_{\rm f}} \left( v_z'-u_z' \right).
  \label{eq:mom_p_2D_O1}
\end{align}

For modes of the form $A' \propto \exp(\omega t - ik_x - ik_z)$, we
find that the system only has non-zero solutions when the
determinant is zero,
\begin{multline}
  \left( \omega -i v_z k_z \right)
  \left( \omega -i v_z k_z +  t_{\rm f}^{-1} \right)  \\
  \left(\omega^2
  + \left( (1+\epsilon)t_{\rm f}^{-1} -i k_z v_{z,0} \right) \omega - \epsilon
  t_{\rm f}^{-1}ik_z v_{z,0} 
  \right) =0\,.
  \label{eq:inset_det}
\end{multline}
The first term represents a pure particle mode falling at the terminal
velocity, while the second term represents particle motion damped by gas drag.
The last factor of this expression has the solutions
\begin{align}
  \omega
  &= \frac{v_z k_z}{2} i
  + \frac{(1+\epsilon)t_{\rm f}^{-1} }{2} 
  \left( -1 \pm \sqrt{1 - \frac{k_z^2 v_{z,0}^2 t_{\rm f}^2}{(1+\epsilon)^2} 
    +2\frac{(\epsilon-1)k_z v_{z,0}t_{\rm f}}{(1+\epsilon)^2}i } \right).
  \label{eq:sol_2D}
\end{align}
The real part of the square root term is always below unity 
 (${\rm Re} \left( \sqrt{1-ix} \right)<1$ for any $x$), so the perturbation
is damped.

In summary, this analysis shows that there are no growing modes under the
assumption of incompressibility. 
Possibly, unstable modes could be found when relaxing the assumption of
incompressibility, or by using more realistic equations of state or by
exploring non-local perturbation techniques. We leave this for future work,
given the complexity of such investigations.

\section{Toy model dispersion relation}
\label{sec:toy}

\begin{figure}[t!]
  \centering
  \includegraphics{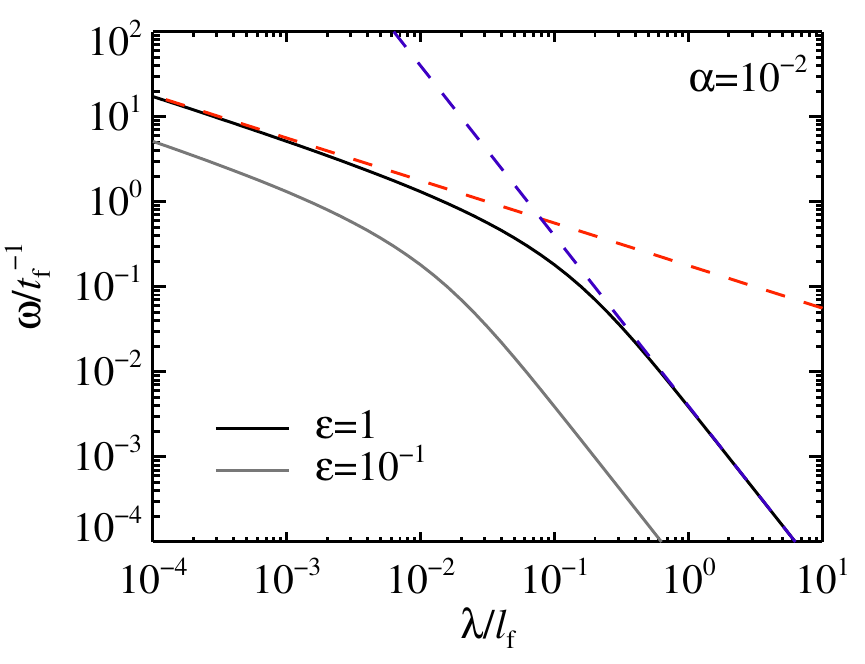}\\
  \caption{
  Linear behaviour of the toy model. 
  In black, the real solution to the dispersion relation is given
  (Eq.\,\ref{eq:disp_simple}, with $\alpha=0.01$). The knee, the largest scale
  for fast growth, is located at $\lambda_{\rm knee} \approx 0.08$.
  The gray curve is the result for a dust-to-gas ratio of $0.1$ as opposed to
  unity.
  The dashed red line gives the high $k$ approximation
  (Eq.\,\ref{eq:disp_simple_simple}) and
  the dashed blue line gives the low $k$ approximation
  (Eq.\,\ref{eq:large_scale_limit}). 
  } 
  \label{fig:growthrate}
\end{figure}

We start with making the ansatz that 
\begin{align}
  u = \alpha (\epsilon - \epsilon_0) v\,,
  \label{eq:ans}
\end{align}
which removes the explicit dependency on the equations for the gas component.
Here $\alpha$ is a proportionality parameter that encapsulates the viscosity
dependency and remains to be determined through numerical simulations\footnote{
Alternatively, one could assume a more general functional dependency of the
form $u(v,\epsilon)$, similar to \citet{Chiang_2010}. Then the friction term
can be linearised to the form $-\frac{1}{t_{\rm f}}(v + v' -u
-\frac{\partial u}{\partial v} v' -\frac{\partial u}{\partial \epsilon} \epsilon')$. 
In order to reduce to expression \ref{eq:ans}, we have to assume
$\frac{\partial u}{\partial v}$ is zero around equilibrium. 
Then, the expression $\alpha v$ corresponds to $\frac{\partial u}{\partial
\epsilon}$.
}.
We also, for simplicity, assume a constant gas density, $\rho = \rho_0$.
Subsequently, the drag term in the particle momentum equation takes the form
\begin{align}
  - \frac{1}{t_{\rm f}} \left( v-u \right) 
  = - \frac{v}{t_{\rm f}} \left[ 1-\alpha (\epsilon - \epsilon_0) \right]\,.
\end{align}

In equilibrium, the dust-to-gas ratio, $\epsilon = \rho_{\rm p}/\rho$, is
constant. 
The momentum equation then shows that particles move, as desired, with terminal
velocity,
\begin{align}
  v_0 &= - gt_{\rm f}\,.
\end{align}

We now linearize the system of particle equations
\begin{align}
  \partial_t \rho_{\rm p}' 
  + \rho_{\rm p,0} \partial_z v' + v_0 \partial_z \rho_{\rm p}' &= 0 \,, \\
  \partial_t v' + v_0  \partial_z v' 
  &= -g - \frac{v_0+v'}{t_{\rm f}} (1 - \alpha \epsilon')\,.
\end{align}
The last two terms simplify\footnote{
When the gas density is not constant the expansion of
$\epsilon = \rho_{\rm p}/\rho = $ goes as
$\epsilon + \epsilon ' = \rho_{\rm p}/\rho + (1/\rho) \rho_{\rm p}' -
(\rho_{\rm p}/\rho^2)\rho'
= \epsilon + \rho_{\rm p}'/\rho - \epsilon  \rho'/\rho$. 
Then the validity of the model relies on the last term of the expansion to be
small.
}
to 
\begin{align}
  -g - \frac{v_0+v'}{t_{\rm f}} (1 - \alpha \epsilon') 
&=  - \frac{v'}{t_{\rm f}} - g\alpha \frac{\rho_{\rm p}'}{\rho_0}\,.
\end{align}

Taking now modes of the form $A' \propto \exp \left( \omega t - i k z \right)$
we are left with the following system of equations
\begin{align}
\begin{pmatrix}
  \omega -ikv_0 & -ik\rho_{\rm p,0} \\ 
  \alpha \frac{g}{\rho_0}  & \omega - ikv_0 + \frac{1}{t_{\rm f}}
\end{pmatrix}
\begin{pmatrix}
  \rho_{\rm p}' \\  v'
\end{pmatrix}
=
\begin{pmatrix}
  0 \\ 0
\end{pmatrix}
.
\end{align}
Non-zero solutions are found when 
\begin{align}
  \beta^2 + \frac{\beta}{t_{\rm f}} + i\alpha \epsilon_0 g k =0,
\end{align}
where $\beta = \omega - ikv_0$. 
Thus we find
\begin{align}
  \beta = \frac{1}{2t_{\rm f}} \left( -1 \pm \sqrt{1-4\alpha \epsilon_0 g
  t_{\rm f}^2 ki} \right) 
\end{align}
where the last term has a positive real part larger than $1$, for any product
$\alpha \epsilon_0 g t_{\rm f}^2$ different from $0$. This reproduces
Eq.\,(\ref{eq:disp_simple}), allowing the approximation of the two limit cases,
Eq.\,(\ref{eq:disp_simple_simple})
at small scales and Eq.\,(\ref{eq:large_scale_limit}) at large scales.
The shape of the dispersion relation and the two limit cases can be inspected
in Fig.\,\ref{fig:growthrate}.

\section{Particle number test}
\label{ap:np}

\begin{figure}[t!]
  \centering
  \includegraphics{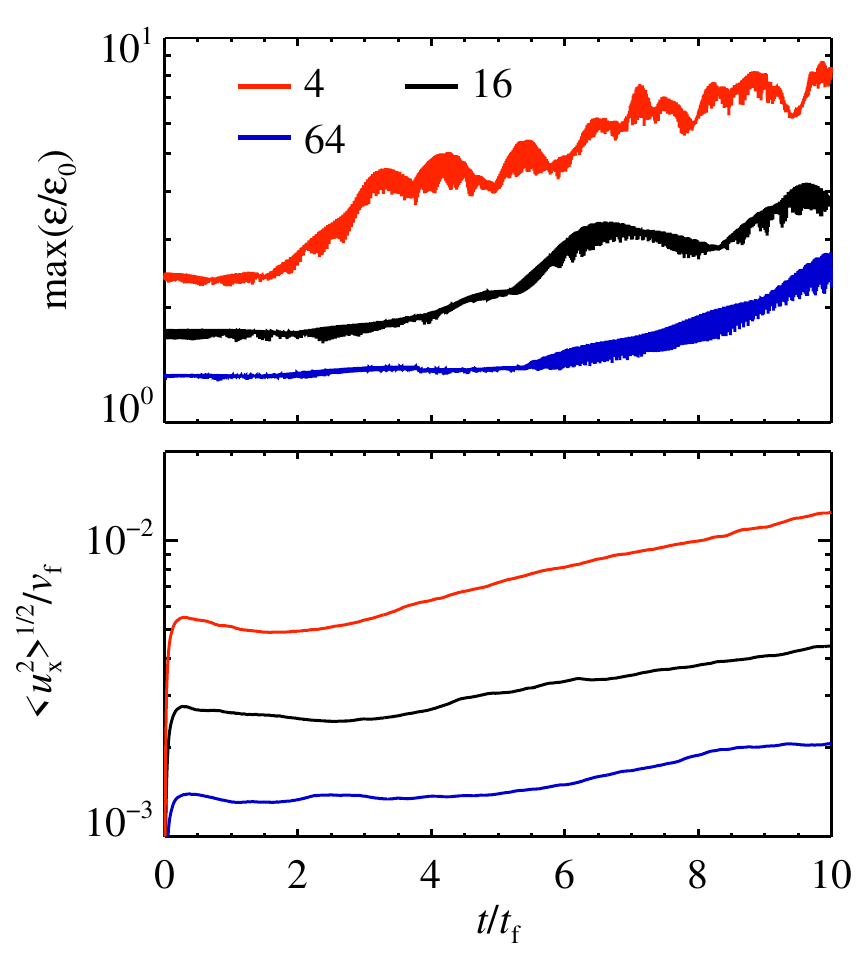}\\
  \caption{
  Evolution of the maximal particle density (\emph{top}) and the horizontal
  gas dispersion (\emph{bottom}), for 4, 16, and 32 particles per gridcell
  (respectively red, black, and blue curves). 
  The initial maximal particle overdensity is reduced from $\epsilon/\epsilon_0
  = 2.4$ to $1.6$ and finally $1.3$, when increasing the particle number by a
  factor of $4$ each time. 
  The lowest particle number simulation shows fastest growth. In higher
  particle number runs the initial dormant phase persists longer and
  growth rates become lower.
  Results from {\texttt run2.n4, run2, run2.n64}.
  } 
  \label{fig:t_np}
\end{figure}

Particle numbers of 16 superparticles per gridcell are sufficient to capture
correctly the evolution of the particle--gas mixture. However, increased
particle numbers decrease the noise amplitude that is initially injected.
In Fig.\,\ref{fig:t_np} we show the evolution of the maximal particle density
and gas velocity dispersion, as function of the particle number.
Because of the non-linear nature of the drafting instability one can see that
the decreased noise amplitude with increased particle number prolongs a dormant
state before the instability comes fully into effect and growth rates decrease
moderately. 
However, if one ignores the protracted dormant phase, growth rates between $16$
(our nominal value) and $64$ particles per gridcells are undistinguishable,
although slower than the $4$ particles per gridcell case.

\bibliographystyle{aa}        
\bibliography{references}     

\end{document}